\RequirePackage{lineno}
\documentclass[aps,prd,twocolumn,showpacs,superscriptaddress,longbibliography,floatfix]{revtex4-1}  % for review and submission
\usepackage{graphicx,epsfig}  % needed for figures
\usepackage{dcolumn}   % needed for some tables
\usepackage{bm}        % for math
\usepackage{amssymb}   % for math
\usepackage{amsmath}   % for math

\def\GeV  {\ensuremath{\mathrm{ Ge\kern -0.1em V } }}
\def\GeVc2{\ensuremath{\mathrm{ Ge\kern -0.1em V }\kern -0.2em /c^2 }}
\def\MeVc2{\ensuremath{\mathrm{ Me\kern -0.1em V }\kern -0.2em /c^2 }}

\newcommand\invisiblesection[1]{%
  \refstepcounter{section}%
  \addcontentsline{toc}{section}{\protect\numberline{\thesection}#1}%
  \sectionmark{#1}}
\begin{document}
%==============================================================================

\widetext
% the following line is for submission, including submission to the arXiv!!
\noindent%
\mbox{FERMILAB-PUB-18-015-E\hspace{35mm}{\em Published in Phys. Rev. D as DOI: 10.1103/PhysRevD.97.112007}}
%\hspace{5.2in} \mbox{FERMILAB-PUB-18-015-E}

\title{Tevatron Run II combination of the effective leptonic electroweak mixing angle}
% Last update: 
\affiliation{Institute of Physics, Academia Sinica, Taipei, Taiwan 11529, Republic of China}
\affiliation{Argonne National Laboratory, Argonne, Illinois 60439, USA}
\affiliation{University of Athens, 157 71 Athens, Greece}
\affiliation{Institut de Fisica d'Altes Energies, ICREA, Universitat Autonoma de Barcelona, E-08193, Bellaterra (Barcelona), Spain}
\affiliation{Baylor University, Waco, Texas 76798, USA}
\affiliation{Istituto Nazionale di Fisica Nucleare Bologna, \ensuremath{^{bbb}}University of Bologna, I-40127 Bologna, Italy}
\affiliation{University of California, Davis, Davis, California 95616, USA}
\affiliation{University of California, Los Angeles, Los Angeles, California 90024, USA}
\affiliation{Instituto de Fisica de Cantabria, CSIC-University of Cantabria, 39005 Santander, Spain}
\affiliation{Carnegie Mellon University, Pittsburgh, Pennsylvania 15213, USA}
\affiliation{Enrico Fermi Institute, University of Chicago, Chicago, Illinois 60637, USA}
\affiliation{Comenius University, 842 48 Bratislava, Slovakia; Institute of Experimental Physics, 040 01 Kosice, Slovakia}
\affiliation{Joint Institute for Nuclear Research, RU-141980 Dubna, Russia}
\affiliation{Duke University, Durham, North Carolina 27708, USA}
\affiliation{Fermi National Accelerator Laboratory, Batavia, Illinois 60510, USA}
\affiliation{University of Florida, Gainesville, Florida 32611, USA}
\affiliation{Laboratori Nazionali di Frascati, Istituto Nazionale di Fisica Nucleare, I-00044 Frascati, Italy}
\affiliation{University of Geneva, CH-1211 Geneva 4, Switzerland}
\affiliation{Glasgow University, Glasgow G12 8QQ, United Kingdom}
\affiliation{Harvard University, Cambridge, Massachusetts 02138, USA}
\affiliation{Division of High Energy Physics, Department of Physics, University of Helsinki, FIN-00014, Helsinki, Finland; Helsinki Institute of Physics, FIN-00014, Helsinki, Finland}
\affiliation{University of Illinois, Urbana, Illinois 61801, USA}
\affiliation{The Johns Hopkins University, Baltimore, Maryland 21218, USA}
\affiliation{Institut f\"{u}r Experimentelle Kernphysik, Karlsruhe Institute of Technology, D-76131 Karlsruhe, Germany}
\affiliation{Center for High Energy Physics: Kyungpook National University, Daegu 702-701, Korea; Seoul National University, Seoul 151-742, Korea; Sungkyunkwan University, Suwon 440-746, Korea; Korea Institute of Science and Technology Information, Daejeon 305-806, Korea; Chonnam National University, Gwangju 500-757, Korea; Chonbuk National University, Jeonju 561-756, Korea; Ewha Womans University, Seoul, 120-750, Korea}
\affiliation{Ernest Orlando Lawrence Berkeley National Laboratory, Berkeley, California 94720, USA}
\affiliation{University of Liverpool, Liverpool L69 7ZE, United Kingdom}
\affiliation{University College London, London WC1E 6BT, United Kingdom}
\affiliation{Centro de Investigaciones Energeticas Medioambientales y Tecnologicas, E-28040 Madrid, Spain}
\affiliation{Massachusetts Institute of Technology, Cambridge, Massachusetts 02139, USA}
\affiliation{University of Michigan, Ann Arbor, Michigan 48109, USA}
\affiliation{Michigan State University, East Lansing, Michigan 48824, USA}
\affiliation{Institution for Theoretical and Experimental Physics, ITEP, Moscow 117259, Russia}
\affiliation{University of New Mexico, Albuquerque, New Mexico 87131, USA}
\affiliation{The Ohio State University, Columbus, Ohio 43210, USA}
\affiliation{Okayama University, Okayama 700-8530, Japan}
\affiliation{Osaka City University, Osaka 558-8585, Japan}
\affiliation{University of Oxford, Oxford OX1 3RH, United Kingdom}
\affiliation{Istituto Nazionale di Fisica Nucleare, Sezione di Padova, \ensuremath{^{ccc}}University of Padova, I-35131 Padova, Italy}
\affiliation{University of Pennsylvania, Philadelphia, Pennsylvania 19104, USA}
\affiliation{Istituto Nazionale di Fisica Nucleare Pisa, \ensuremath{^{ddd}}University of Pisa, \ensuremath{^{eee}}University of Siena, \ensuremath{^{fff}}Scuola Normale Superiore, I-56127 Pisa, Italy, \ensuremath{^{ggg}}INFN Pavia, I-27100 Pavia, Italy, \ensuremath{^{hhh}}University of Pavia, I-27100 Pavia, Italy}
\affiliation{University of Pittsburgh, Pittsburgh, Pennsylvania 15260, USA}
\affiliation{Purdue University, West Lafayette, Indiana 47907, USA}
\affiliation{University of Rochester, Rochester, New York 14627, USA}
\affiliation{The Rockefeller University, New York, New York 10065, USA}
\affiliation{Istituto Nazionale di Fisica Nucleare, Sezione di Roma 1, \ensuremath{^{iii}}Sapienza Universit\`{a} di Roma, I-00185 Roma, Italy}
\affiliation{Mitchell Institute for Fundamental Physics and Astronomy, Texas A\&M University, College Station, Texas 77843, USA}
\affiliation{Istituto Nazionale di Fisica Nucleare Trieste, \ensuremath{^{jjj}}Gruppo Collegato di Udine, \ensuremath{^{kkk}}University of Udine, I-33100 Udine, Italy, \ensuremath{^{lll}}University of Trieste, I-34127 Trieste, Italy}
\affiliation{University of Tsukuba, Tsukuba, Ibaraki 305, Japan}
\affiliation{Tufts University, Medford, Massachusetts 02155, USA}
\affiliation{Waseda University, Tokyo 169, Japan}
\affiliation{Wayne State University, Detroit, Michigan 48201, USA}
\affiliation{University of Wisconsin-Madison, Madison, Wisconsin 53706, USA}
\affiliation{Yale University, New Haven, Connecticut 06520, USA}
\affiliation{LAFEX, Centro Brasileiro de Pesquisas F\'{i}sicas, Rio de Janeiro, RJ 22290, Brazil}
\affiliation{Universidade do Estado do Rio de Janeiro, Rio de Janeiro, RJ 20550, Brazil}
\affiliation{Universidade Federal do ABC, Santo Andr\'{e}, SP 09210, Brazil}
\affiliation{University of Science and Technology of China, Hefei 230026, People's Republic of China}
\affiliation{Universidad de los Andes, Bogot\'{a}, 111711, Colombia}
\affiliation{Charles University, Faculty of Mathematics and Physics, Center for Particle Physics, 116 36 Prague 1, Czech Republic}
\affiliation{Czech Technical University in Prague, 116 36 Prague 6, Czech Republic}
\affiliation{Institute of Physics, Academy of Sciences of the Czech Republic, 182 21 Prague, Czech Republic}
\affiliation{Universidad San Francisco de Quito, Quito 170157, Ecuador}
\affiliation{LPC, Universit\'{e} Blaise Pascal, CNRS/IN2P3, Clermont, F-63178 Aubi\`ere Cedex, France}
\affiliation{LPSC, Universit\'{e} Joseph Fourier Grenoble 1, CNRS/IN2P3, Institut National Polytechnique de Grenoble, F-38026 Grenoble Cedex, France}
\affiliation{CPPM, Aix-Marseille Universit\'{e}, CNRS/IN2P3, F-13288 Marseille Cedex 09, France}
\affiliation{LAL, Univ. Paris-Sud, CNRS/IN2P3, Universit\'{e} Paris-Saclay, F-91898 Orsay Cedex, France}
\affiliation{LPNHE, Universit\'{e}s Paris VI and VII, CNRS/IN2P3, F-75005 Paris, France}
\affiliation{CEA Saclay, Irfu, SPP, F-91191 Gif-Sur-Yvette Cedex, France}
\affiliation{IPHC, Universit\'{e} de Strasbourg, CNRS/IN2P3, F-67037 Strasbourg, France}
\affiliation{IPNL, Universit\'{e} Lyon 1, CNRS/IN2P3, F-69622 Villeurbanne Cedex, France and Universit\'{e} de Lyon, F-69361 Lyon CEDEX 07, France}
\affiliation{III. Physikalisches Institut A, RWTH Aachen University, 52056 Aachen, Germany}
\affiliation{Physikalisches Institut, Universit\"{a}t Freiburg, 79085 Freiburg, Germany}
\affiliation{II. Physikalisches Institut, Georg-August-Universit\"{a}t G\"{o}ttingen, 37073 G\"{o}ttingen, Germany}
\affiliation{Institut f\"{u}r Physik, Universit\"{a}t Mainz, 55099 Mainz, Germany}
\affiliation{Ludwig-Maximilians-Universit\"{a}t M\"{u}nchen, 80539 M\"{u}nchen, Germany}
\affiliation{Panjab University, Chandigarh 160014, India}
\affiliation{Delhi University, Delhi-110 007, India}
\affiliation{Tata Institute of Fundamental Research, Mumbai-400 005, India}
\affiliation{University College Dublin, Dublin 4, Ireland}
\affiliation{Korea Detector Laboratory, Korea University, Seoul, 02841, Korea}
\affiliation{CINVESTAV, Mexico City 07360, Mexico}
\affiliation{Nikhef, Science Park, 1098 XG Amsterdam, the Netherlands}
\affiliation{Radboud University Nijmegen, 6525 AJ Nijmegen, the Netherlands}
\affiliation{Joint Institute for Nuclear Research, RU-141980 Dubna, Russia}
\affiliation{Institution for Theoretical and Experimental Physics, ITEP, Moscow 117259, Russia}
\affiliation{Moscow State University, Moscow 119991, Russia}
\affiliation{Institute for High Energy Physics, Protvino, Moscow region 142281, Russia}
\affiliation{Petersburg Nuclear Physics Institute, St. Petersburg 188300, Russia}
\affiliation{Instituci\'{o} Catalana de Recerca i Estudis Avan\c{c}ats (ICREA) and Institut de F\'{i}sica d'Altes Energies (IFAE), 08193 Bellaterra (Barcelona), Spain}
\affiliation{Uppsala University, 751 05 Uppsala, Sweden}
\affiliation{Taras Shevchenko National University of Kyiv, Kiev, 01601, Ukraine}
\affiliation{Lancaster University, Lancaster LA1 4YB, United Kingdom}
\affiliation{Imperial College London, London SW7 2AZ, United Kingdom}
\affiliation{The University of Manchester, Manchester M13 9PL, United Kingdom}
\affiliation{University of Arizona, Tucson, Arizona 85721, USA}
\affiliation{University of California Riverside, Riverside, California 92521, USA}
\affiliation{Florida State University, Tallahassee, Florida 32306, USA}
\affiliation{Fermi National Accelerator Laboratory, Batavia, Illinois 60510, USA}
\affiliation{University of Illinois at Chicago, Chicago, Illinois 60607, USA}
\affiliation{Northern Illinois University, DeKalb, Illinois 60115, USA}
\affiliation{Northwestern University, Evanston, Illinois 60208, USA}
\affiliation{Indiana University, Bloomington, Indiana 47405, USA}
\affiliation{Purdue University Calumet, Hammond, Indiana 46323, USA}
\affiliation{University of Notre Dame, Notre Dame, Indiana 46556, USA}
\affiliation{Iowa State University, Ames, Iowa 50011, USA}
\affiliation{University of Kansas, Lawrence, Kansas 66045, USA}
\affiliation{Louisiana Tech University, Ruston, Louisiana 71272, USA}
\affiliation{Northeastern University, Boston, Massachusetts 02115, USA}
\affiliation{University of Michigan, Ann Arbor, Michigan 48109, USA}
\affiliation{Michigan State University, East Lansing, Michigan 48824, USA}
\affiliation{University of Mississippi, University, Mississippi 38677, USA}
\affiliation{University of Nebraska, Lincoln, Nebraska 68588, USA}
\affiliation{Rutgers University, Piscataway, New Jersey 08855, USA}
\affiliation{Princeton University, Princeton, New Jersey 08544, USA}
\affiliation{State University of New York, Buffalo, New York 14260, USA}
\affiliation{University of Rochester, Rochester, New York 14627, USA}
\affiliation{State University of New York, Stony Brook, New York 11794, USA}
\affiliation{Brookhaven National Laboratory, Upton, New York 11973, USA}
\affiliation{Langston University, Langston, Oklahoma 73050, USA}
\affiliation{University of Oklahoma, Norman, Oklahoma 73019, USA}
\affiliation{Oklahoma State University, Stillwater, Oklahoma 74078, USA}
\affiliation{Oregon State University, Corvallis, Oregon 97331, USA}
\affiliation{Brown University, Providence, Rhode Island 02912, USA}
\affiliation{University of Texas, Arlington, Texas 76019, USA}
\affiliation{Southern Methodist University, Dallas, Texas 75275, USA}
\affiliation{Rice University, Houston, Texas 77005, USA}
\affiliation{University of Virginia, Charlottesville, Virginia 22904, USA}
\affiliation{University of Washington, Seattle, Washington 98195, USA}

\author{T.~Aaltonen \ensuremath{^{\dagger}}}
\affiliation{Division of High Energy Physics, Department of Physics, University of Helsinki, FIN-00014, Helsinki, Finland; Helsinki Institute of Physics, FIN-00014, Helsinki, Finland}
\author{V.M.~Abazov \ensuremath{^{\ddagger}}}
\affiliation{Joint Institute for Nuclear Research, RU-141980 Dubna, Russia}
\author{B.~Abbott \ensuremath{^{\ddagger}}}
\affiliation{University of Oklahoma, Norman, Oklahoma 73019, USA}
\author{B.S.~Acharya \ensuremath{^{\ddagger}}}
\affiliation{Tata Institute of Fundamental Research, Mumbai-400 005, India}
\author{M.~Adams \ensuremath{^{\ddagger}}}
\affiliation{University of Illinois at Chicago, Chicago, Illinois 60607, USA}
\author{T.~Adams \ensuremath{^{\ddagger}}}
\affiliation{Florida State University, Tallahassee, Florida 32306, USA}
\author{J.P.~Agnew \ensuremath{^{\ddagger}}}
\affiliation{The University of Manchester, Manchester M13 9PL, United Kingdom}
\author{G.D.~Alexeev \ensuremath{^{\ddagger}}}
\affiliation{Joint Institute for Nuclear Research, RU-141980 Dubna, Russia}
\author{G.~Alkhazov \ensuremath{^{\ddagger}}}
\affiliation{Petersburg Nuclear Physics Institute, St. Petersburg 188300, Russia}
\author{A.~Alton \ensuremath{^{\ddagger}}\ensuremath{^{kk}}}
\affiliation{University of Michigan, Ann Arbor, Michigan 48109, USA}
% ...

\collaboration{CDF Collaboration}
\altaffiliation[With visitors from]{
\ensuremath{^{a}}University of British Columbia, Vancouver, BC V6T 1Z1, Canada,
\ensuremath{^{b}}Istituto Nazionale di Fisica Nucleare, Sezione di Cagliari, 09042 Monserrato (Cagliari), Italy,
\ensuremath{^{c}}University of California Irvine, Irvine, CA 92697, USA,
\ensuremath{^{d}}Institute of Physics, Academy of Sciences of the Czech Republic, 182 21, Czech Republic,
\ensuremath{^{e}}CERN, CH-1211 Geneva, Switzerland,
\ensuremath{^{f}}Cornell University, Ithaca, NY 14853, USA,
\ensuremath{^{g}}University of Cyprus, Nicosia CY-1678, Cyprus,
\ensuremath{^{h}}Office of Science, U.S. Department of Energy, Washington, DC 20585, USA,
\ensuremath{^{i}}University College Dublin, Dublin 4, Ireland,
\ensuremath{^{j}}ETH, 8092 Z\"{u}rich, Switzerland,
\ensuremath{^{k}}University of Fukui, Fukui City, Fukui Prefecture, Japan 910-0017,
\ensuremath{^{l}}Universidad Iberoamericana, Lomas de Santa Fe, M\'{e}xico, C.P. 01219, Distrito Federal,
\ensuremath{^{m}}University of Iowa, Iowa City, IA 52242, USA,
\ensuremath{^{n}}Kinki University, Higashi-Osaka City, Japan 577-8502,
\ensuremath{^{o}}Kansas State University, Manhattan, KS 66506, USA,
\ensuremath{^{p}}Brookhaven National Laboratory, Upton, NY 11973, USA,
\ensuremath{^{q}}Istituto Nazionale di Fisica Nucleare, Sezione di Lecce, Via Arnesano, I-73100 Lecce, Italy,
\ensuremath{^{r}}Queen Mary, University of London, London, E1 4NS, United Kingdom,
\ensuremath{^{s}}University of Melbourne, Victoria 3010, Australia,
\ensuremath{^{t}}Muons, Inc., Batavia, IL 60510, USA,
\ensuremath{^{u}}Nagasaki Institute of Applied Science, Nagasaki 851-0193, Japan,
\ensuremath{^{v}}National Research Nuclear University, Moscow 115409, Russia,
\ensuremath{^{w}}Northwestern University, Evanston, IL 60208, USA,
\ensuremath{^{x}}University of Notre Dame, Notre Dame, IN 46556, USA,
\ensuremath{^{y}}Universidad de Oviedo, E-33007 Oviedo, Spain,
\ensuremath{^{z}}CNRS-IN2P3, Paris, F-75205 France,
\ensuremath{^{aa}}Universidad Tecnica Federico Santa Maria, 110v Valparaiso, Chile,
\ensuremath{^{bb}}Sejong University, Seoul 143-747, Korea,
\ensuremath{^{cc}}The University of Jordan, Amman 11942, Jordan,
\ensuremath{^{dd}}Universite catholique de Louvain, 1348 Louvain-La-Neuve, Belgium,
\ensuremath{^{ee}}University of Z\"{u}rich, 8006 Z\"{u}rich, Switzerland,
\ensuremath{^{ff}}Massachusetts General Hospital, Boston, MA 02114 USA,
\ensuremath{^{gg}}Harvard Medical School, Boston, MA 02114 USA,
\ensuremath{^{hh}}Hampton University, Hampton, VA 23668, USA,
\ensuremath{^{ii}}Los Alamos National Laboratory, Los Alamos, NM 87544, USA,
\ensuremath{^{jj}}Universit\`{a} degli Studi di Napoli Federico II, I-80138 Napoli, Italy
}
\noaffiliation
\collaboration{D0 Collaboration}
\altaffiliation[With visitors from]{
\ensuremath{^{kk}}Augustana University, Sioux Falls, SD 57197, USA,
\ensuremath{^{ll}}The University of Liverpool, Liverpool L69 3BX, UK,
\ensuremath{^{mm}}Deutsches Elektronen-Synchrotron (DESY), Notkestrase 85, Germany,
\ensuremath{^{nn}}Consejo Nacional de Ciencia y Tecnologia (Conacyt), M-03940 Mexico City, Mexico,
\ensuremath{^{oo}}SLAC, Menlo Park, CA 94025, USA,
\ensuremath{^{pp}}University College London, London WC1E 6BT, UK,
\ensuremath{^{qq}}Centro de Investigacion en Computacion - IPN, CP 07738 Mexico City, Mexico,
\ensuremath{^{rr}}Universidade Estadual Paulista, S\~{a}o Paulo, SP 01140, Brazil,
\ensuremath{^{ss}}Karlsruher Institut f\"{u}r Technologie (KIT) - Steinbuch Centre for Computing (SCC), D-76128 Karlsruher, Germany,
\ensuremath{^{tt}}Office of Science, U.S. Department of Energy, Washington, D.C. 20585, USA,
\ensuremath{^{uu}}American Association for the Advancement of Science, Washington, D.C. 20005, USA,
\ensuremath{^{vv}}National Academy of Science of Ukraine (NASU) - Kiev Institute for Nuclear Research (KINR), Kyiv 03680, Ukraine,
\ensuremath{^{ww}}University of Maryland, College Park, MD 20742, USA,
\ensuremath{^{xx}}European Organization for Nuclear Research (CERN), CH-1211 Gen\'{e}ve 23, Switzerland,
\ensuremath{^{yy}}Purdue University, West Lafayette, IN 47907, USA,
\ensuremath{^{zz}}Institute of Physics, Belgrade, CS-11080 Belgrade, Serbia,
\ensuremath{^{aaa}}P.N. Lebedev Physical Institute of the Russian Academy of Sciences, 119991, Moscow, Russia
}
\noaffiliation
% Last update: 
            % CDF and D0 authors
                             % (includes institutions and visitors)
\date{January 18, 2018}

\begin{abstract}
  Drell-Yan lepton pairs produced in the process
  $p \bar{p} \rightarrow \ell^+\ell^- + X$ through an intermediate
  $\gamma^*/Z$ boson have an asymmetry in their angular distribution
  related to the spontaneous symmetry breaking of the
  electroweak force and the associated mixing of its neutral gauge
  bosons. The CDF and D0 experiments have measured the effective-leptonic 
  electroweak mixing parameter $\sin^2\theta^{\rm lept}_{\rm eff}$ using
  electron and muon pairs selected from the full
  Tevatron proton-antiproton data sets collected in 2001-2011,
  corresponding to \mbox{9-10}~fb$^{-1}$ of integrated luminosity.
  The combination of these measurements yields the most precise
  result from hadron colliders,
  $\sin^2 \theta^{\rm lept}_{\rm eff} = 0.23148 \pm 0.00033$.
  This result is consistent with, and approaches in precision, the best
  measurements from electron-positron colliders. The standard model
  inference of the on-shell electroweak mixing parameter $\sin^2\theta_W$,
  or equivalently the $W$-boson mass $M_W$, using the \textsc{zfitter}
  software package yields
  $\sin^2 \theta_W  =  0.22324 \pm 0.00033$ or equivalently,
  $M_W  =  80.367 \pm 0.017 \;{\rm GeV}/c^2$.
\end{abstract}

\maketitle

\vfill

%\setcounter{page}{2}

%\linenumbers

%==============================================================================
\section{\label{sec:intro}
Introduction}

At the Fermilab Tevatron proton-antiproton ($p\bar{p}$) collider,
Drell-Yan~\cite{DrellYan} lepton ($\ell$) pairs are produced in the process
$p \bar{p} \rightarrow \ell^+\ell^- + X$ through an intermediate
$\gamma^*/Z$ boson, where $X$ represents inclusively any other collision
products. The forward-backward asymmetry in the polar-angle distribution
of the $\ell^-$ in the Collins-Soper (CS) frame \cite{CollinsSoperFrame} as a
function of the $\ell^+\ell^-$-pair invariant mass is directly sensitive
to the effective-leptonic electroweak mixing parameter
$\sin^2\theta^{\rm lept}_{\rm eff}$. The effective-leptonic parameter
is measured using electron and muon pairs ($\ell = e$ and $\mu$).
The electroweak-mixing parameter $\sin^2\theta_W$ \cite{PDGreviews} is obtained
indirectly in the context of standard model (SM) calculations with the following
input parameters: the fine structure constant, the Fermi constant, the strong
interaction coupling constant, and the masses of the top quark, $Z$ boson,
and Higgs boson. In this SM context, $\sin^2\theta_W$ and the
$W$-boson mass are related, and a comparison of the $W$-boson mass inferred
from $\sin^2\theta_W$ to the directly measured mass tests the consistency
of the SM. Such tests require
precision measurements of $\sin^2\theta^{\rm lept}_{\rm eff}$, and
results from hadron colliders such as the Tevatron are complementary
to those from electron-positron colliders.
\par
The Drell-Yan process and the production of quark pairs in
high-energy $e^+e^-$ collisions are analogous processes:
   $q\bar{q} \rightarrow \ell^+\ell^-$ and
   $e^+e^-   \rightarrow q\bar{q}$.
The $\sin^2\theta^{\rm lept}_{\rm eff}$ parameter of processes
involving leptons has been investigated at the
LEP-1 and SLC~\cite{LEPfinalZ,LEPfinalZ2} $e^+e^-$ colliders
operating on or in the vicinity of the $Z$-boson pole mass,
and at the Tevatron
\cite{CDFIIsw2e,zA4ee21prd,cdfAfb9mmprd,cdfAfb9eeprd,
      D0sw2e1,D0sw2e5,D0sw2e10,D0sw2m10}
and LHC \cite{ATLASsw2eff,CMSsw2eff1,LHCBsw2eff} hadron colliders.
Investigations at hadron colliders use Drell-Yan pairs whose range
of invariant masses about the $Z$-boson resonant peak is broad
relative to the  $e^+e^-$ collider investigations.
The mixing parameter has been accurately measured at the LEP-1 and SLC
colliders, where processes with leptons in the final state are also used.
The combined average of six measurements from these lepton colliders
\cite{*[{The measurements are $A_{\rm fb}^{0,\ell}$, ${\cal A}_{\ell}(P_{\tau})$,
   ${\cal A}_{\ell}({\rm SLD})$, $A_{\rm fb}^{0,{\rm b}}$,
   $A_{\rm fb}^{0,{\rm c}}$, and $Q_{\rm fb}^{{\rm had}}$ in }] [{.}]
   LEPfinalZ}  
yields a value of $0.23149 \pm 0.00016$ \cite{LEPfinalZ2}.
However, a
3.2 standard-deviation difference exists between the two most precise
individual measurements. The combined measurement of the $b$-quark
forward-backward asymmetry ($A_{\rm FB}^{0,{\rm b}})$ 
with the LEP-1 detectors yields
$\sin^2\theta^{\rm lept}_{\rm eff} = 0.23221 \pm 0.00029$,
while the SLD left-right polarization asymmetry of $Z$-boson
production $({\cal A}_\ell)$
yields $\sin^2\theta^{\rm lept}_{\rm eff} = 0.23098 \pm 0.00026$.
This provides a strong motivation for an accurate determination
of $\sin^2\theta^{\rm lept}_{\rm eff}$
by the Tevatron experiments.

\subsection{\label{ssec:ewkCouplings}
Electroweak couplings}

The production of Drell-Yan lepton pairs at
the Born level proceeds through two parton-level processes,
\begin{eqnarray}
  q\bar{q} \rightarrow \gamma^* \rightarrow \ell^+\ell^- \;\; {\rm and} \;\; q\bar{q} \rightarrow Z \rightarrow \ell^+\ell^- ,  
\label{eqnBornProcess}
\end{eqnarray}
where the $q$ and $\bar{q}$ are a quark and antiquark, respectively, that 
originate from the colliding hadrons. The virtual photon couples the vector
currents of the incoming and outgoing fermions $(f)$, and the
spacetime structure of the photon-fermion interaction vertex
may be represented as
$\langle \bar{f} | Q_f \gamma_\mu |f\rangle$,
where $Q_f$, the strength of the coupling, is the fermion charge
(in units of $e$), and $|f\rangle$ is the spinor for fermion $f$.
The interaction vertex of a fermion with a $Z$ boson contains both
vector $(V)$ and axial-vector $(A)$ current components, and its
structure is
$\langle \bar{f} | g_V^f \gamma_\mu + g_A^f \gamma_\mu\gamma_5 |f\rangle$.
The Born-level coupling strengths are
\begin{eqnarray}
  g_V^f = T_3^f - 2Q_f \: \sin^2\theta_W \;\; {\rm and} \;\; g_A^f = T_3^f ,
\label{eqnBornCoupling}
\end{eqnarray}
where $T_3^f$ is the third component of the fermion weak isospin,
which is $T_3^f = \frac{1}{2}$ $(-\frac{1}{2})$
for positively (negatively) charged fermions.
Radiative corrections alter the
Born-level couplings into {\rm effective} couplings.
At the Born level in the SM, and in all orders of the on-shell
renormalization scheme~\cite{OnShellScheme}, the $\sin^2\theta_W$
parameter is related to the $W$-boson mass $M_W$ and the
$Z$-boson mass $M_Z$ by $\sin^2\theta_W =  1 - M_W^2/M_Z^2$.
Since the $Z$-boson mass is accurately known
(to $\pm 0.0021$ GeV/$c^2$ \cite{LEPfinalZ,LEPfinalZ2}),
the inference of the on-shell $\sin^2 \theta_W$ is equivalent to an
indirect $W$-boson mass measurement.
The angular distributions of $\ell^+\ell^-$ pairs in the final state
of the Drell-Yan process and of $\ell^+\ell^-$ or $q\bar{q}$ pairs
in the final state of $e^+e^-$ collisions are sensitive to the
effective $\sin^2 \theta_W$ parameter at the lepton vertex,
$\sin^2\theta^{\rm lept}_{\rm eff}$.

\subsection{\label{ssec:fbAsymmetry}
The forward-backward asymmetry}

The rapidity, transverse
momentum, and mass of a particle or a system of particles are
represented by $y$, $p_{\rm T}$, and $M$, respectively. The energy and
momentum of particles are represented as $E$ and $\vec{p}$, respectively.
In the laboratory frame, the  $p\bar{p}$ collision axis is the $z_{\rm lab}$
axis, with the positive direction defined to be along the direction of the
proton. The transverse component of any vector, such as the momentum vector,
is defined relative to that axis. The rapidity is
$y = \frac{1}{2} \, \ln[\,(E + p_{\rm z})/(E - p_{\rm z})\,]$,
where $p_{\rm z}$ is the component of the momentum vector along the
$z_{\rm lab}$ axis.
\par
\begin{figure}
\includegraphics
   [scale=0.4]
   {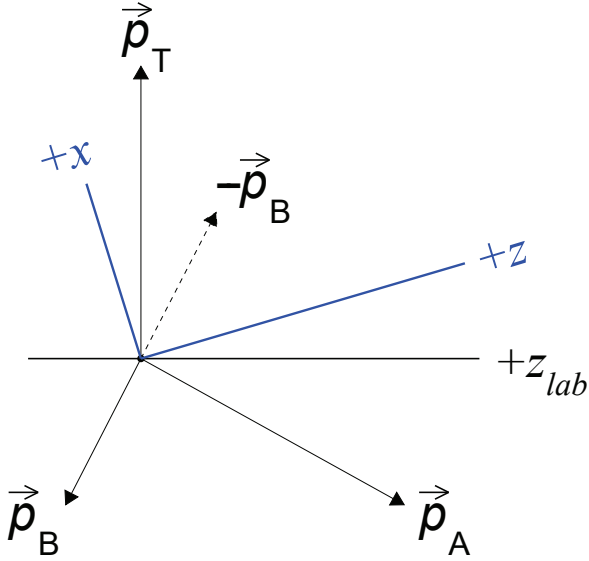}
\caption{\label{fig_CSframe}
Representation of the Collins-Soper coordinate axes $(x,z)$ in the
lepton-pair rest frame relative to the laboratory $z$ axis ($z_{\rm lab}$).
These three axes are in the plane formed by the
proton ($\vec{p}_{\rm A}$) and antiproton ($\vec{p}_{\rm B}$)
momentum vectors in the lepton-pair rest frame. The $z$ axis is the
angular bisector of $\vec{p}_{\rm A}$ and $-\vec{p}_{\rm B}$.
The $y$ axis is along the direction of
$\vec{p}_{\rm B} \times \vec{p}_{\rm A}$, and the $x$ axis
is in the direction opposite to the transverse component of
$\vec{p}_{\rm A}+\vec{p}_{\rm B}$. 
}
\end{figure}

 
The angular kinematic properties of leptons from the Drell-Yan process
are defined in the rest frame of the exchanged $\gamma^*/Z$ boson.
The $\ell^-$ direction is chosen to define the polar and azimuthal
angles of the lepton pair, which are denoted as $\vartheta$ and $\varphi$,
respectively. The ideal positive $z$ axis coincides with the
direction of the incoming quark so that the definition of $\vartheta$
parallels the definition used in $e^+e^-$ collisions at
LEP~\cite{LEPfinalZ,LEPfinalZ2}.
This frame is approximated by the CS rest
frame~\cite{CollinsSoperFrame} for $p\bar{p}$ collisions, depicted
in Fig.~\ref{fig_CSframe}.
\par
The CS frame angle $\vartheta$ is reconstructed
using the following laboratory-frame quantities: the lepton energies,
the lepton momenta along the beam line, the
dilepton invariant mass, $M$, and the dilepton transverse momentum, 
$p_{\rm T}$.
The polar angle of the negatively charged lepton is calculated from
\begin{eqnarray}
  \cos \vartheta = \frac{ l^-_+l^+_- - l^-_-l^+_+ }
                 { M \sqrt{M^2 + p_{\rm T}^2} }  \; ,
\label{eqnCStheta}
\end{eqnarray}
where $l_\pm = (E \pm p_{\rm z})$ and
the $+$ $(-)$ superscript specify that $l_\pm$ is for
the positively (negatively) charged lepton. Similarly, the
CS expression for $\varphi$ in terms of
laboratory-frame quantities is given by
\begin{eqnarray}
  \tan \varphi = \frac{\sqrt{M^2 + p_{\rm T}^2}}{M} \;
	\frac{\vec{\Delta} \cdot \widehat{R}_{\rm T}}
	     {\vec{\Delta} \cdot \widehat{p}_{\rm T}} \: ,
\label{eqnCSphi}
\end{eqnarray}
where $\vec{\Delta}$ is the difference between the $\ell^-$ and
$\ell^+$ laboratory-frame momentum vectors; $\widehat{R}_{\rm T}$ is the
unit vector along $\vec{p}_p \times \vec{p}$, with
$\vec{p}_p$ being the proton momentum vector and $\vec{p}$ the
lepton-pair momentum vector; and $\widehat{p}_{\rm T}$ is the
unit vector along the transverse component of the lepton-pair
momentum vector. At $p_{\rm T} = 0$, the angular distribution
is azimuthally symmetric. The right-hand sides of the definitions
of $\cos \vartheta$ and $\tan \varphi$ are invariant
under Lorentz boosts along the $z$ direction in the laboratory frame.
\par
The angular distribution of Drell-Yan lepton pairs is defined
as the ratio of the production cross section to the
angle-integrated production cross section. Its general structure
consists of terms derived from nine helicity cross sections that
describe the polarization state of the boson
\cite{MirkesA0to7a, MirkesA0to7b},
\begin{eqnarray}
\frac{dN}{d\Omega}
  & \propto &
        \: (1 + \cos^2 \vartheta) +  \nonumber \\
  &   & A_0 \:\frac{1}{2} \:
             (1 -3\cos^2 \vartheta) + \nonumber \\
  &   & A_1 \: \sin 2\vartheta
               \cos \varphi +   \nonumber \\
  &   & A_2 \: \frac{1}{2} \:
               \sin^2 \vartheta
               \cos 2\varphi +  \nonumber \\
  &   & A_3 \: \sin \vartheta
               \cos \varphi +   \nonumber \\
  &   & A_4 \: \cos \vartheta + \nonumber \\
  &   & A_5 \: \sin^2 \vartheta
               \sin 2\varphi +  \nonumber \\
  &   & A_6 \: \sin 2\vartheta
               \sin \varphi +   \nonumber \\
  &   & A_7 \: \sin \vartheta
               \sin \varphi \: .
\label{eqnAngDistr}
\end{eqnarray}
The coefficients $A_{0-7}$ are
functions of kinematic variables of the boson and vanish when the
lepton-pair transverse momentum approaches zero, except for $A_4$,
which contributes to the tree-level amplitude and generates the
forward-backward
asymmetry in $\cos \vartheta$. Thus, at zero transverse
momentum, the angular distribution reduces to the tree-level
form $1 + \cos^2 \vartheta + A_4\cos \vartheta$. In the CS frame, 
the $A_0$, $A_2$, and $A_4$ coefficients are large relative 
to the other coefficients.

The $A_4 \cos\vartheta$ term violates parity conservation, and is due to
the interference of the amplitudes of the vector and axial-vector
currents. Its presence induces an asymmetry in the
$\varphi$-integrated $\cos \vartheta$ dependence of the cross section.
Two sources contribute: the interference between
the $Z$-boson vector and axial-vector amplitudes, and the
interference between the photon vector and $Z$-boson axial-vector
amplitudes. The asymmetry component from the $\gamma^*$-$Z$
interference cross section depends on axial-vector couplings
$g_A^f$ to fermions $f$ that are independent of $\sin^2 \theta_W$.
The asymmetry component from $Z$-boson self-interference depends on
a product of $g_V^\ell$ and $g_V^q$ from the lepton and quark
vertices, and thus is related to $\sin^2 \theta_W$.
At the Born level, this product is
\begin{eqnarray}
   T_3^\ell \: (1 - 4|Q_\ell|\sin^2\theta_W) \;
   T_3^q    \: (1 - 4|Q_q|\sin^2\theta_W) \:,
\end{eqnarray}
where $\ell$ and $q$ denote the lepton and quark, respectively.
For the Drell-Yan process, the relevant quarks are predominantly
the light quarks $u$, $d$, and $s$. The coupling factor has an
enhanced sensitivity to $\sin^2\theta_W$ at the lepton-$Z$
vertex: for a $\sin^2\theta_W$ value of $0.223$, a 1\% variation
in $\sin^2\theta_W$ changes the lepton factor
$(1 - 4|Q_\ell|\sin^2\theta_W)$
by about 8\%, and it changes the quark factor
$(1 - 4|Q_q|\sin^2\theta_W)$
by about 1.5\% (0.4\%) for the $u$ ($d$ or $s$) quark.
Electroweak radiative corrections do not significantly alter
this Born-level interpretation. Loop and vertex
electroweak radiative corrections induce multiplicative
form-factor corrections \cite{Dizet, zfitter621, zfitter642} to
the $T_3^f$ and $\sin^2\theta_W$ terms
that change their values by a few percent
\cite{zA4ee21prd}.

\begin{figure}
\includegraphics
   [scale=0.4]
   {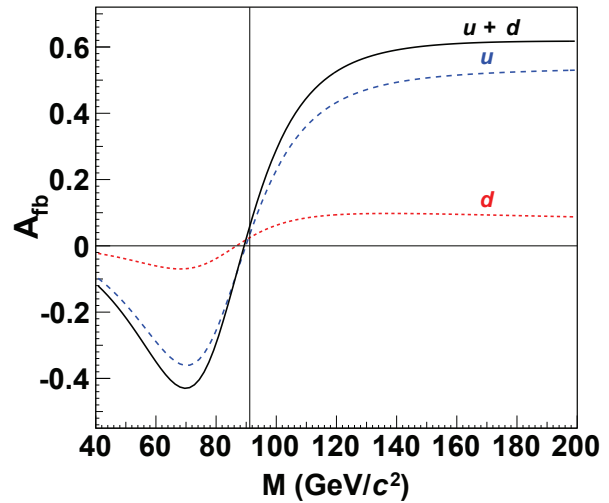}
\caption{\label{fig_loAfbVSmass}
 Typical dependence of $A_{\rm fb}$ as a function of the lepton-pair
 invariant mass $M$. The vertical line is at $M = M_Z$. The label
 $u$ + $d$ denotes the overall asymmetry, and the labels $u$ and $d$
 denote the contribution to the overall asymmetry from quarks with
 charge $2/3$ and $-1/3$, respectively. The asymmetry identified by
 the $u$ or $d$ label is defined as
 $(\sigma_{\rm f}^q - \sigma_{\rm b}^q)/\sigma$, where $q = u$ or $d$,
 $\sigma_{\rm f(b)}$ is the forward (backward) cross section, and
 $\sigma$ is the corresponding cross section from quarks of all charges.
 Thus, the overall asymmetry is the sum of the asymmetries identified
 by the $u$ and $d$ labels.
}
\end{figure}

The forward-backward asymmetry of the polar-angle distribution is
defined as
\begin{eqnarray}
  A_{\rm fb}(M) = \frac{\sigma_{\rm f}(M) - \sigma_{\rm b}(M)}
                       {\sigma_{\rm f}(M) + \sigma_{\rm b}(M)} 
                = \frac{3}{8}A_4(M) \:,
\label{eqnAfbDef}
\end{eqnarray}
where $\sigma_{\rm f}$ is the Drell-Yan cross section for the
forward (f) orientation of lepton pairs,
$\cos \vartheta \geq 0$, and $\sigma_{\rm b}$ is for the
backward (b) orientation of lepton pairs, $\cos \vartheta < 0$. 
Figure~\ref{fig_loAfbVSmass} shows the typical dependence of the
asymmetry as a function of the lepton-pair invariant mass from a
Drell-Yan quantum chromodynamics calculation.
The offset of $A_{\rm fb}$ from zero at $M = M_Z$ is related
to $\sin^2\theta_W$. At lepton-pair masses much smaller or larger
than the $Z$-pole mass, the asymmetry is dominated by the component
from $\gamma^*$-$Z$ interference, whose cross section contains the
factor $(M^2-M_Z^2)/M^2$ from the photon and $Z$-boson
propagators, the factor for the flux of quarks from the parton
distribution functions (PDF) of the proton, and the factor for the
coupling term $4/9$ ($1/9$) at the quark-photon vertex for the
charge $2/3$ ($-1/3$) quark.
Consequently, the asymmetry distribution is sensitive to both
$\sin^2\theta_W$ and the PDF of the proton.

\subsection{\label{ssec:measMethod}
Extraction of $\sin^2\theta^{\rm lept}_{\rm eff}$}

The $\sin^2\theta^{\rm lept}_{\rm eff}$ parameter is derived
from comparisons of the measurement of $A_{\rm fb}(M)$ in bins of mass
covering the measurement range,
and predictions of the measurement (templates) for various
input values of the effective-leptonic parameter. The value that
minimizes the $\chi^2$ between the measurement and templates is the
extracted value of $\sin^2\theta^{\rm lept}_{\rm eff}$.
Detector effects that bias the asymmetry measurement
are removed from the measurement by CDF, and incorporated
into the templates by D0. In both the CDF and D0 analyses,
the forward-backward asymmetries of electron and muon pairs are
separately measured.
\par
The CDF and D0 measurements are based on multiple analyses with
improvements over time. Different PDFs have been utilized by
CDF and D0, as well as slightly different electroweak
correction techniques based on the SM in the extraction of the
$\sin^2\theta^{\rm lept}_{\rm eff}$ parameter. For the
combination of electron- and muon-channel results within an
experiment, and for the Tevatron combination, a framework
with two common components, denoted as the ``common framework'',
is adopted to obtain consistent results for the effective-leptonic
parameter.
The common components are electroweak-radiative corrections
calculated using the SM software package
\textsc{zfitter} 6.43 \cite{Dizet, zfitter621, zfitter642},
which is used with LEP-1 and SLD measurement inputs for precision
tests of the SM at the $Z$ pole \cite{LEPfinalZ,LEPfinalZ2}, and
NNPDF 3.0~\cite{ nnpdf301, nnpdf302,  nnpdf303, nnpdf304,
                 nnpdf305, nnpdf306,  nnpdf307}
PDFs, which include new HERA and LHC data, and replace the
Tevatron $W$-asymmetry measurements with lepton-asymmetry
measurements from the LHC.
A summary of the CDF and D0 measurements is presented in
Sec.~\ref{sec:TheMeasurements}, along with the strategies used to
achieve results consistent with the common-framework agreement.
The combination of these CDF and D0 results is presented in
Sec.~\ref{sec:d0cdfCombination}, and
Sec.~\ref{sec:theEndSummary} provides the summary.
 
%==============================================================================

\section{\label{sec:TheMeasurements}
Input measurements}

The Fermilab Tevatron collider provides $p\bar{p}$ collisions at a
center-of-momentum energy of 1.96~TeV. 
The CDF \cite{detCDFII,*refCOT,*refSVXII,*refCEM,*refChad,
	      *refPEM,*refPHA,*refPES,*refCMU,*cdfR2CLC,*refTrigXFT,
              *refTrigSVT1,*refTrigSVT2,*refTrigSVTUpgr}
and D0 \cite{detD0RunI,*detD0L2MUON,*detD0RunIItrig,*detD0L3DAQ,*detD0muon,                   *detD0RunII,*detD0L1CAL,*detD0L0,*detD0SMT,*detD0muonID,                        *detD0emID} 
detectors are general-purpose 
detectors that surround the Tevatron collision regions. For both 
detectors, precision charged-particle tracking detectors (trackers) 
immersed in solenoidal magnetic fields are located
around the beam line. Beyond the trackers are projective-tower
electromagnetic and hadronic calorimeters with transverse and
longitudinal segmentation,
which are used for the identification of electrons, photons, and 
clusters of outgoing particles (jets),
and for measurements of their kinematic properties. The outer regions of
each detector consist of hadron absorbers and trackers for the detection
and identification of muons.
\par
For particle trajectories, the
polar and azimuthal angles of the coordinate systems are denoted by
$\theta_{\rm lab}$ and $\phi_{\rm lab}$ respectively.
The pseudorapidity of a particle is $\eta = -\ln\tan(\theta_{\rm lab}/2)$.
Detector coordinates are specified as $(\eta_{\rm det}, \phi_{\rm det})$,
where $\eta_{\rm det}$ is the pseudorapidity relative to the detector
center $(z_{\rm lab}=0)$. For particle energies measured in the
calorimeters, the transverse energy $E_{\rm T}$ is defined as
$E_{\rm T} = E \sin\theta_{\rm lab}$.
\par
Both the CDF and D0 measurements utilize electron and muon pairs
reconstructed in high-$p_{\rm T}$ electron and muon data samples,
respectively. During data taking, events
with high-$p_{\rm T}$ electrons and muons are selected online by
triggers. Offline selection criteria are applied to these samples to
improve the quality of the sample used for the asymmetry measurements.

\subsection{\label{ssec:CDFsummary}
CDF measurement}

The CDF measurements for the electron \cite{cdfAfb9eeprd} and
muon \cite{cdfAfb9mmprd} channels use the full Tevatron
Run II data set, corresponding to an
integrated luminosity of 9~fb$^{-1}$ of $p\bar{p}$ collisions.
Similar methods are used for both analyses. The $A_{\rm fb}$ 
measurements are corrected for detector effects. The effects of detector 
resolution and quantum electrodynamics (QED) final-state radiation (FSR)
are removed from the measurements using the CDF simulation of
Drell-Yan events. The templates are strictly quantum chromodynamics
(QCD) calculations of $A_{\rm fb}$.

Approximately 485~000 electron pairs and 277~000 muon pairs
are utilized in the measurements. Details of the electron- and
muon-selection criteria are presented in
Refs.~\cite{cdfAfb9eeprd} and \cite{cdfAfb9mmprd}, respectively.
Kinematic, fiducial, and lepton-identification criteria are
applied. As Drell-Yan leptons are usually produced in isolation
from the other activity in the event, isolation criteria that
limit event activity around the leptons are also applied.
In the electron channel, one electron must be
detected in the central-calorimeter region
$0.05 < |\eta_{\rm det}| < 1.05$, and its partner electron
can be either in the central-calorimeter region or the
plug-calorimeter region $1.2 < |\eta_{\rm det}| < 2.8$. For
central-central electron pairs, one electron is required to have
$E_{\rm T} > 25$ GeV and the other to have
$E_{\rm T} > 15$ GeV. For central-plug electron pairs, both
electrons are required to have $E_{\rm T} > 20$ GeV. 
In the muon channel, the selection requires both muons to
be predominantly within
the central muon-detector region $|\eta_{\rm det}|<1$, and have
$p_{\rm T} > 20$ GeV/$c$.
For the asymmetry measurements, electron and muon pairs are selected
to have invariant masses in the range \mbox{50-350}~GeV/$c^2$ and
\mbox{50-1000}~GeV/$c^2$, respectively. The upper limit for muon
pairs is larger to accommodate the significant resolution
smearing of the muon momentum at high masses.
\par
The $A_{\rm{fb}}$ measurement uses a data-driven event-weighting
method \cite{evtwtAFBmethod}, which is equivalent to performing
individual measurements of the asymmetry in $|\cos\vartheta|$
bins and then combining them. The standard expression for the
asymmetry in a bin is
\begin{eqnarray}
  A_{\rm fb} = \frac{ N_{\rm f}/(\epsilon A)_{\rm f} - N_{\rm b}/(\epsilon A)_{\rm b} }
                    { N_{\rm f}/(\epsilon A)_{\rm f} + N_{\rm b}/(\epsilon A)_{\rm b} }
               \: ,
\label{eqnAfbExp}
\end{eqnarray}
where $N_{\rm f(b)}$ and $(\epsilon A)_{\rm f(b)}$ are the signal
event counts, and the efficiency $(\epsilon)$ and acceptance
$(A)$ product, respectively, of forward (backward) lepton pairs.
Only leptons reconstructed within the central region,
where tracking performance is optimal with small charge
misidentification probability, are used to determine the
forward (backward) orientation of a pair. For these leptons,
the track-finding efficiency is 99\% \cite{refCdfTrackingEff}. 
Within a localized $|\cos\vartheta|$ bin, the forward and
backward dependence of the acceptance and efficiency of its
lepton pairs cancels out to first order so that
\begin{eqnarray}
  A_{\rm fb} \approxeq \frac{ N_{\rm f} - N_{\rm b} }
                            { N_{\rm f} + N_{\rm b} } \:.
\label{eqnAfbExpCancel}
\end{eqnarray}
The similarity of $(\epsilon A)_{\rm f}$ and $(\epsilon A)_{\rm b}$
for a value of $|\cos\vartheta|$ is related to the proximity of
lepton-pair trajectories within the detector to those of
the ones with the charges exchanged. With the exchange, the magnitude
of $\cos\vartheta$ is unchanged but the sign is reversed, \mbox{i.e.},
forwards and backwards are interchanged. The exchange transforms a
lepton with charge $Q$ into one with charge $-Q$ but leaves its
laboratory momentum $\vec{p}$ unchanged. For a typical lepton
with $p_{\rm T} \sim 40$ GeV/$c$, the radius of curvature in the
magnetic-field-bend plane is 100~m. For reference, the radial
extent of the precision trackers in this plane is 1.3~m. Thus,
the trajectories of a lepton and its charge-exchanged configuration
through the detector are localized, so to first order, their
probabilities of being reconstructed and selected are the same.
\par
The expected angular dependencies of the numerator-event
difference and denominator-event sum of
Eq.~(\ref{eqnAfbExpCancel}) are derived using
Eq.~(\ref{eqnAngDistr}).
The numerator difference is proportional
to $2 A_4 |\cos \vartheta|$, and the denominator sum
to $2 (1 + \cos^2 \vartheta + \cdots)$, where
$1 + \cos^2 \vartheta + \cdots$ denotes the $\cos \vartheta$
symmetric terms from Eq.~(\ref{eqnAngDistr}). Together they
yield $A_{\rm fb} = A_4 \xi$, where
$\xi = |\cos \vartheta| / (1 + \cos^2 \vartheta + \cdots)$.
Each bin is an independent measurement of $A_4$, with an
uncertainty of $\sigma / \xi$, where $\sigma$ is the statistical
uncertainty of $A_{\rm fb}$. When the measurements are combined,
each bin has a statistical weight of $\sigma^{-2}$ and a sampling
weight of $\xi^2$. The binned measurements are reformulated into
an unbinned-asymmetry expression of the form shown in the
right-hand side of Eq.~(\ref{eqnAfbExpCancel}) using event weights.
Weights for individual events in the numerator and denominator
remove the angular dependencies of the event difference and sum,
respectively, and provide the sampling weight for the combination
of events across $|\cos\vartheta|$.
\par
Because the event-weighting method needs events for
corrections, kinematic regions with few or no events
are eliminated from the acceptance region of
the measurement and template calculations. Consequently,
the kinematic-acceptance region of the electron-pair is
restricted to $|y|<1.7$, and that of the muon-pair to $|y|<1$.
Small secondary effects not corrected by the event-weighting
method are removed with the simulation.
\par
The simulation of Drell-Yan events, described below, uses
\textsc{pythia}~6.2 \cite{Pythia621} with CTEQ5L \cite{Cteq5pdf} PDFs
to generate events and includes QED FSR for the decay leptons. To
account for QED FSR from promptly decaying hadrons and their decay
products, the event generation then uses           
\textsc{photos} 2.0~\cite{Photos20a, Photos20b, Photos20c}. This is
followed by the detector simulation based on
\textsc{geant}-3 and \textsc{gflash}~\cite{nimGflash}.
Some of the kinematic distributions of the generated $\gamma^*/Z$
boson are adjusted with event weights for better agreement between
the simulation and the data.
For the electron channel, simulation adjustments cover all
aspects of boson production because of the extended boson-rapidity
coverage of the measurement. 
The generated boson kinematics are adjusted using the data and
the \textsc{resbos}~\cite{ResBos1, ResBos2, ResBos3, ResBosc221}
calculation with CTEQ6.6 \cite{cteq66pdf} PDFs.
The generator-level
$p_{\rm T}$ distribution of the boson is adjusted so that the
reconstructed $p_{\rm T}$ distribution of simulated events
matches the distribution of the data in two rapidity bins,
$0 < |y| < 0.8$ and
$|y| \ge 0.8$. The generator-level boson-mass distribution is
adjusted with a mass-dependent factor, which is the
ratio of the \textsc{resbos} boson-mass distribution calculated
using CTEQ6.6 PDFs relative to the
\textsc{pythia}~6.4 \cite{pythia64}
distribution calculated using CTEQ5L PDFs.
For the muon channel, the generator-level $p_{\rm T}$ distribution
of the boson is adjusted so that the reconstructed $p_{\rm T}$
distribution of simulated events matches the data.
\par
The energy scales of both the data and simulation are calibrated
to a common standard following Ref.~\cite{muPcorrMethod}. The standard 
consists of idealized calibration distributions derived from the same 
generated-event samples of the simulation using a detector with perfect
calibrations but with the observed resolutions. The energy resolution
in the simulation is calibrated to that of the data.
In addition to the generator level tuning, other distributions
such as the time-dependent Tevatron beam-luminosity profile and
detector responses near boundaries are tuned. Response
adjustments are consistently applied to parity-symmetric variables
such as $|\eta_{\rm det}|$ or $|\cos\vartheta|$. They are needed
for an unbiased detector-resolution unfolding of the
asymmetry distribution in mass and $\cos \vartheta$ for the data.
Unfolding matrices are used, and the unfolding statistically removes
the effects of resolution smearing and QED FSR. The simulation
is also used to derive the error matrix for the $A_{\rm{fb}}$
measurement.
\par
The backgrounds are from the production of QCD dijets, $W+\rm{jets}$,
$\gamma^*/Z \rightarrow \tau\tau$, diboson
({\it WW}, {\it WZ}, and {\it ZZ}), and $t\bar{t}$ events.
QCD dijet backgrounds
are estimated using the data. Other backgrounds listed above
are estimated with \textsc{pythia}~6.2 \cite{Pythia621}.
For the electron channel, the overall background level amounts
to 1.1\% over the mass range of the asymmetry measurement.
For the muon channel, the overall  background level amounts to
0.5\% over the mass range of the asymmetry measurement.
All backgrounds are subtracted from the data.
\par
The CDF templates for $A_{\rm{fb}}$ which are compliant with the
specifications of the common framework are denoted as
``common-framework compliant templates''. They are calculated using the
\textsc{powheg-box} next-to-leading-order (NLO) implementation
\cite{Powheg-Box} of the Drell-Yan process \cite{PowhegBoxVBP}
followed by \textsc{pythia}~6.41 \cite{pythia64}
parton-showering. The combined implementation has
next-to-leading-log resummation accuracy. The NNPDF 3.0
next-to-next-to-leading order (NNLO) PDFs are used for the
parton fluxes.
\par
The complex-valued \textsc{zfitter}
form factors are incorporated into the \textsc{powheg-box}
amplitudes as specified in the Appendix. The QED photon propagator
correction from fermion loops is also included.
The implementation of these form
factors provides an enhanced Born approximation (EBA) to the
electroweak couplings. For consistency with the \textsc{zfitter}
calculations, the NNPDFs selected are derived assuming a value of
the strong-interaction coupling of 0.118 at the $Z$-boson mass.
With \textsc{zfitter} corrections, the electroweak mixing parameter
for the templates is the static on-shell $\sin^2\theta_W$.
The asymmetry is directly sensitive to the effective mixing terms,
which are provided by \textsc{zfitter} as
$\kappa_f \sin^2\theta_W$, where $\kappa_f$ denotes a
fermion-flavor $(f)$ dependent form factor. Unlike the directly
observable effective mixing terms, $\sin^2\theta_W$ and $\kappa_f$
are inferred in the context of the SM and their inputs are specified 
in the Appendix. The effective mixing terms are functions of 
$\sin^2\theta_W$ and a mass scale. For comparisons with other 
measurements, the value of the $\sin^2 \theta^{\rm lept}_{\rm eff}$ 
parameter is defined at the $Z$ pole and its value is
$\operatorname{Re}[\kappa_e(\sin^2 \theta_W,M_Z^2)] \sin^2 \theta_W$.
\par
The NNPDF 3.0 parton distributions consist of an
ensemble of 100 equally probable PDFs for which the value of a
calculation is the average of the calculations over the ensemble,
and the rms about the average is the PDF uncertainty. A measurement
can be incorporated into the ensemble of a PDF fit without
regenerating the ensemble via various Bayesian methods that reweight
each ensemble PDF \cite{BayesPdfRewt}. With the Giele-Keller (GK)
method, the ensemble PDFs, numbered 1 to $N$, are reweighted with
the likelihood of the prediction using the ensemble PDF relative to
the measurement,
\begin{eqnarray}
   w_k = \frac{ \exp(-{\frac{1}{2}\chi^2_k}) }
              {\sum_{l=1}^{N} \exp(-{\frac{1}{2}\chi^2_l}) } \;,
\label{GKwtEqn}
\end{eqnarray}
where $w_k$ is the weight for PDF number $k$, and $\chi^2_k$ is
the $\chi^2$ between the measurement and the prediction using
that PDF \cite{GKpdfRewt,pdfGKweightMethod}.
\par
For the extraction of the $\sin^2\theta^{\rm lept}_{\rm eff}$
parameter, common-framework compliant templates with varying values of the
effective-leptonic parameter are calculated for each of the ensemble
PDFs of NNPDF 3.0, and the GK-weighting method is used to evaluate the
ensemble average and rms of the effective-leptonic parameter derived
from the measurement and templates. Consequently, additional PDF
constraints from the asymmetry measurement are incorporated into
the ensemble of PDFs.
\par
The electron-channel measurement of the asymmetry, along with
common-framework compliant templates and the error matrix of the measurement,
are used to extract the $\sin^2\theta^{\rm lept}_{\rm eff}$
parameter. The result for the electron channel is
\begin{eqnarray}
  \sin^2\theta^{\rm lept}_{\rm eff} 
     & = & 0.23248  \pm \: 0.00049  \: {\rm (stat)}  \nonumber \\
     &   & \pm \: 0.00004  \: {\rm (syst)}  \nonumber \\
     &   & \pm \: 0.00019  \: {\rm (PDF)}.
\label{eqnCDFechan}
\end{eqnarray}
The systematic uncertainty consists of contributions from the
energy scale and resolution, the backgrounds, and the QCD
scale.
\par
The muon-channel measurement of the asymmetry
and the error matrix of the measurement used for
measurement-to-template $\chi^2$ comparisons are
presented in Ref.~\cite{cdfAfb9mmprd}.
The published result for the muon channel is
\begin{eqnarray}
  \sin^2\theta^{\rm lept}_{\rm eff} 
     & = & 0.2315 \pm \: 0.0009 \: {\rm (stat)}  \nonumber \\
     &   & \pm \: 0.0002 \: {\rm (syst)}  \nonumber \\
     &   & \pm \: 0.0004 \: {\rm (PDF)}.
\label{eqnCDFmuchan}
\end{eqnarray}
The systematic uncertainty consists of contributions from the
energy scale and resolution, the backgrounds, and the 
higher-order terms of QCD. The $A_{\rm{fb}}$ templates used for
the derivation of this result are from previous iterations of the
analysis and not compliant with the specifications of the
common framework. They are calculated with a
modified version of \textsc{resbos} using CTEQ6.6 PDFs and
the \textsc{zfitter} form factors. Systematic uncertainties
are estimated using the \textsc{powheg-box} NLO parton generator
with CT10 NLO PDFs \cite{ct10pdfs}, and followed by
\textsc{pythia}~6.41 parton showering. The PDF uncertainty
is derived from the CT10 uncertainty PDFs at 68\% \mbox{C.L.}
The result shown in Eq.~(\ref{eqnCDFmuchan}) is the standalone
result of the muon channel, but the methods used do not
facilitate a straightforward combination with the electron-channel
analysis.
\par
For the combination of the electron- and muon-channel results, the
measured asymmetries are directly utilized. Both the electron- and
muon-channel measurements are compared against common-framework 
compliant templates for the extraction of the 
$\sin^2 \theta^{\rm lept}_{\rm eff}$ parameter. For comparison purposes, 
the extracted value of the effective-leptonic parameter from the 
muon-asymmetry measurement in this framework is $0.23141 \pm 0.00086$, 
where the uncertainty is statistical only \cite{cdfAfb9eeprd}.
\par
The CDF result combining electron and muon channels presented in
Ref. \cite{cdfAfb9eeprd} is
\begin{eqnarray}
  \sin^2\theta^{\rm lept}_{\rm eff} 
     & = & 0.23221 \pm \: 0.00043 \: {\rm (stat)}  \nonumber \\
     &   & \pm \: 0.00007 \: {\rm (syst)}  \nonumber \\
     &   & \pm \: 0.00016 \: {\rm (PDF)}.
\label{eqnCDFcomb}
\end{eqnarray}
The systematic uncertainty, consisting of contributions from
the energy scale and resolution, the backgrounds, and the QCD
scale, are summarized in Table~\ref{cdftblSystErrComb}. 
\begin{table}
\caption{\label{cdftblSystErrComb}
Summary of the CDF systematic uncertainties on the electron-
and muon-channel combination of the electroweak mixing parameter
$\sin^2\theta^{\rm lept}_{\rm eff}$. The column
labeled $\delta\sin^2\theta^{\rm lept}_{\rm eff}$ gives the
uncertainty of each source. The values
are from Table~VI of Ref. \cite{cdfAfb9eeprd}.
}
\begin{ruledtabular}
\begin{tabular}{lc}
Source  & $\delta\sin^2\theta^{\rm lept}_{\rm eff}$ \\ \hline
Energy scale and resolution   & $\pm \: 0.00002$  \\
Backgrounds                   & $\pm \: 0.00003$  \\
QCD scale                     & $\pm \: 0.00006$  \\
NNPDF 3.0 PDF                 & $\pm \: 0.00016$  \\
\end{tabular}
\end{ruledtabular}
\end{table}
As the $A_{\rm{fb}}$ templates for the electron-
and muon-channel asymmetries are both calculated with the same
common-framework infrastructure, the electron- and muon-channel
comparison $\chi^2$'s between the data and templates
are combined
into a joint $\chi^2$ for the determination of the best-fit
$\sin^2\theta^{\rm lept}_{\rm eff}$. The joint $\chi^2$ takes
into account correlations between the electron- and
muon-channel asymmetries for each of the ensemble PDFs of
NNPDF~3.0. The GK-weighting method using the joint $\chi^2$
incorporates PDF information from both the electron- and
muon-channel asymmetry measurements into the PDF
ensemble, and thus reduces the PDF uncertainty relative to
the default (equal-weight) ensemble.

\subsection{\label{ssec:D0summary}
D0 measurement}

The D0 measurement of $\sin^2\theta^{\rm lept}_{\rm eff}$ is performed in the 
electron \cite{D0sw2e10} and muon \cite{D0sw2m10} channels.  The 
electron- and muon-channel results use 9.7~fb$^{-1}$ and 8.6~fb$^{-1}$ of 
recorded luminosity, respectively.

The asymmetry $A_{\rm{fb}}$ is measured in the electron channel using events with
at least two electromagnetic (EM) clusters reconstructed in the calorimeter. 
They are required to be in the central calorimeter (CC) or end calorimeter (EC)
with transverse momentum $p_{\rm T}>25$~GeV/$c$. Clusters in the CC must be 
matched to reconstructed tracks. For events with both clusters in the EC, only 
one cluster must be track-matched.
Compared to previous D0 results based on 1.1~fb$^{-1}$ 
and 5.0~fb$^{-1}$ of luminosity \cite{D0sw2e1,D0sw2e5}, the acceptance 
is extended from $|\eta_{\rm det}|<1.0$ to $|\eta_{\rm det}|<1.1$ for CC and from 
$1.5<|\eta_{\rm det}|<2.5$ to $1.5<|\eta_{\rm det}|<3.2$ for EC, and previously 
rejected electrons reconstructed near azimuthal CC module boundaries are 
included. By extending the $\eta_{\rm det}$ and module boundary acceptance, a 
$70\%$ increase is achieved in the number of sample events above what would be 
expected from the increase in luminosity to the full data set, for a 
total of approximately 560~000 electron pairs in the final sample. Events 
are categorized as CC-CC, CC-EC, or EC-EC based on the $\eta_{\rm det}$ regions 
of the two electron candidates. 

Muon-channel events are required to have at least two muon 
candidates reconstructed in the tracking and muon systems, with 
transverse momenta $p_{\rm T}>15$~GeV/$c$. Both muon candidates are required to 
have $|\eta_{\rm det}|<1.8$ with at least one muon within $|\eta_{\rm det}|<1.6$, 
and they must have tracks matched in the tracking and muon systems. Tracks are 
required to have opposite curvature. Events with muons nearly back-to-back are 
removed to reduce cosmic-ray background. The large kinematic acceptance yields
a final sample consisting of approximately 481~000 muon pairs.

Simulated Drell-Yan events are generated using leading-order 
\textsc{pythia}~6.23 \cite{Pythia621} with the NNPDF 2.3 \cite{nnpdf302} PDFs 
for the electron channel and NNPDF 3.0 PDFs for the muon channel, followed by a
\textsc{geant}-3-based simulation \cite{GEANT} of the D0 detector. The 
inner-tracker solenoid and muon system toroid polarities are reversed every 
two weeks on average at D0, enabling 
cross-checks and cancellations of charge-dependent asymmetries.  For example, 
the muon Drell-Yan samples are generated with different polarities of the 
solenoid and toroid magnetic fields in the \textsc{geant}-3 simulation, and used
to model the data corresponding to each combination separately. Data and
simulation samples corresponding to different solenoid and toroid polarities 
are weighted to correspond to equal luminosity exposures for each 
solenoid-toroid polarity combination. This weighted combination provides 
cancellation of asymmetries due to variations in detector response. The same 
reconstruction algorithm is used for the data and simulation. 

Relative to previously reported results, new methods of electron energy and 
muon momentum calibration are developed and applied to both data and 
simulation. In addition to scale factors, offset parameters are applied to the 
electron energy as functions of $\eta_{\rm det}$ and instantaneous 
luminosity. For muon momentum, a scale factor is introduced that is dependent 
on charge, $\eta_{\rm det}$, and solenoid polarity. With these calibration 
methods, the systematic uncertainties due to energy and momentum modeling are 
reduced to negligible levels.  

The backgrounds are from the production of multijets, $W+\rm{jets}$, 
$\gamma^*/Z \rightarrow \tau\tau$, dibosons ({\it WW} and {\it WZ}), and 
$t\bar{t}$ events.  The multijet backgrounds are estimated using the data. The 
$W+\rm{jets}$ events are generated using \textsc{alpgen}~\cite{alpgen} 
interfaced to \textsc{pythia}~6.23 for showering and hadronization. The other 
backgrounds are estimated using the \textsc{pythia}~6.23 simulations. At the 
$Z$-boson mass peak, the overall background level is 0.35\% (0.88\%) with 
respect to the total number of selected dielectron (dimuon) events in the 
data.

The $A_{\rm{fb}}$ templates are calculated using \textsc{pythia}~6.23 with 
NNPDF 2.3 (electron channel) and NNPDF 3.0 (muon channel). They are then 
reweighted to incorporate higher-order QCD effects. The $Z$-boson distribution 
as a function of rapidity and transverse momentum $(y,p_{\rm T})$ is reweighted 
in both variables to match that from \textsc{resbos} \cite{ResBos1, ResBos2, 
ResBos3, ResBosc221} with CTEQ6.6 \cite{cteq66pdf} PDFs. The boson-mass 
distribution is reweighted with a mass-dependent NNLO K-factor \cite{wzprod1}. 
The events are processed by the D0 detector simulation to yield templates that 
include detector resolution effects.

The $A_{\rm{fb}}$ distributions in data are obtained as a function of the 
dilepton invariant mass. For the electron channel, this is done separately for 
CC-CC, CC-EC, and EC-EC event categories. The weak mixing parameter is 
extracted from the background-subtracted $A_{\rm{fb}}$ spectrum in the regions 
$75 < M_{ee} < 115$~GeV/$c^2$ for CC-CC and CC-EC events, 
$81 < M_{ee} < 97$~GeV/$c^2$ for EC-EC events, and 
$74 < M_{\mu\mu} < 110$~GeV/$c^2$ for muon events, by comparing the data to 
simulated $A_{\rm{fb}}$ templates corresponding to different input values of 
$\sin^2\theta_W$.

Combining the weak-mixing-parameter results from the three electron event 
categories gives the electron-channel result
\begin{eqnarray}
  \sin^2\theta^{\rm meas}_{\rm eff} 
     & = & 0.23139 \pm \: 0.00043 \: {\rm (stat)}  \nonumber \\
     &   & \pm \: 0.00008 \: {\rm (syst)}  \nonumber \\
     &   & \pm \: 0.00017 \: {\rm (PDF)}.
\label{eqnD0echan_uncorr}
\end{eqnarray}
The sources of systematic uncertainties include energy calibration,
energy resolution smearing, backgrounds, charge misidentification, and electron 
identification \cite{D0sw2e10}. The largest component is from the electron 
identification (0.00007). 
The PDF uncertainty is obtained using the equally probable ensemble
PDFs of NNPDF 2.3 following the method prescribed by the NNPDF group
\cite{nnpdf302}.

The extraction of the weak mixing parameter from the muon channel $A_{\rm{fb}}$ 
distribution in data gives 
\begin{eqnarray}
  \sin^2\theta^{\rm meas}_{\rm eff} 
     & = & 0.22994 \pm \: 0.00059 \: {\rm (stat)}  \nonumber \\
     &   & \pm \: 0.00005 \: {\rm (syst)}  \nonumber \\
     &   & \pm \: 0.00024 \: {\rm (PDF)}.
\label{eqnD0mchan_uncorr}
\end{eqnarray}
The systematic uncertainties include the following sources: momentum 
calibration, momentum resolution smearing, backgrounds, and muon identification 
\cite{D0sw2m10}. The PDF uncertainty is obtained using the 100 equally probable
ensemble PDFs of NNPDF 3.0 following the method prescribed by the NNPDF group
\cite{nnpdf301}.

%
% Corrections
%

Corrections are applied to the values of $\sin^2 \theta^{\rm meas}_{\rm eff}$ in
Eqs.~(\ref{eqnD0echan_uncorr}) and (\ref{eqnD0mchan_uncorr}) to make them
compliant with the agreed-upon common framework using the CDF EBA 
electroweak-radiative correction implementation, and NNPDF 3.0. 

%
%Weak radiative corrections
%

The $A_{\rm{fb}}$ templates used by D0 for both the electron- and muon-channel
analyses are calculated with \textsc{pythia}, which
uses the same fixed value for the effective mixing terms $\sin^2 \theta_{\rm eff}$ 
of all fermions. The EBA implementation incorporates \textsc{zfitter}
weak-interaction corrections and the fermion-loop correction to the photon 
propagator, both of which are complex valued and mass-scale dependent. The 
effect of using a fixed and constant value for all of the 
effective mixing terms is investigated by setting all weak form factors
to unity so that the effective mixing terms for the templates become 
$\sin^2 \theta^{\rm lept}_{\rm eff}$, where only the real part of the photon 
propagator correction, the running $\alpha_{\rm em}$, is retained. This 
implementation, denoted as nonEBA, is the analog to the \textsc{pythia} 
calculation. The difference,
\begin{eqnarray}
  \Delta \sin^2\theta_{\rm eff}^{\rm lept}({\rm \textsc{zfitter}}) &=&
  \sin^2 \theta^{\rm lept}_{\rm eff}({\rm EBA}) - \nonumber \\
  & & \sin^2 \theta^{\rm lept}_{\rm eff}({\rm nonEBA}) \:,
\label{eqnRadCorr}
\end{eqnarray}
provides the correction to the values of 
$\sin^2 \theta^{\rm lept}_{\rm eff}$ derived using \textsc{pythia} templates, which
is needed to convert them to the values derived using \textsc{zfitter} form 
factors.

The difference is calculated using the combination of the CDF electron- and 
muon-channel $A_{\rm{fb}}$ measurements. Both nonEBA and EBA template 
calculations use NNPDF 3.0 NNLO with $\alpha_s(M_Z) = 0.118$. Each template 
contains about $10^9$ generated events, and the uncertainty in $A_{\rm{fb}}$ for 
the mass bin containing the $Z$-boson mass is about $5\times 10^{-5}$. 
Differences are calculated for 23 ensemble PDFs whose best-fit mixing parameter
value from EBA templates is near the average value derived over all ensemble 
PDFs. The GK-weighted average and rms of 
$\Delta \sin^2\theta_{\rm eff}^{\rm lept}({\rm \textsc{zfitter}})$ over the
23 PDFs are 0.00022 and 0.00002, respectively.
To accommodate the statistical uncertainty of the effective-leptonic parameter, an 
additional uncertainty of 0.00003 is assigned, resulting in a total 
(EBA$-$nonEBA) uncertainty of $\pm \: 0.00004$. This value of 
$+0.00022 \pm 0.00004$ is the correction from the \textsc{pythia} framework, 
which assumes a single value for all the effective mixing parameters, to the 
\textsc{zfitter}-based one. 

%
%\subsection{\label{ssec:pdfChoice}
%PDF choice}
%

The D0 electron-channel $A_{\rm{fb}}$ template calculations use NNPDF 2.3. 
Therefore, Eq.~(\ref{eqnD0echan_uncorr}) must be corrected to comply with the
common framework choice of NNPDF 3.0. To calculate the correction, the value of 
$\sin^2 \theta^{\rm lept}_{\rm eff}$ is extracted using a default NNPDF 3.0 
template and multiple NNPDF 2.3 templates.
A \textsc{pythia} template of $5 \times 10^8$ events is generated, using NNPDF 
3.0 with fixed $\sin^2\theta_W$ input, and taken as pseudodata after applying 
fast-simulation kinematic requirements. In addition, 40 templates of $3 \times 
10^8$ events each are generated, using NNPDF 2.3 with varying $\sin^2\theta_W$ 
inputs and applying the same selection criteria.  These templates are used to 
obtain the best $\chi^2$ fit for the $\sin^2\theta_W$ value.  The difference 
between the input value for NNPDF 3.0 and the value extracted from NNPDF 2.3, 
denoted by $\Delta \sin^2 \theta^{\rm lept}_{\rm eff}({\rm PDF})$, is used to 
provide a correction to the D0 electron-channel value of 
$\sin^2 \theta^{\rm lept}_{\rm eff}$ derived using NNPDF 2.3 to that derived using 
NNPDF 3.0. The value of $\Delta \sin^2 \theta^{\rm lept}_{\rm eff}({\rm PDF})$ 
based on the D0 analysis is $-0.00024 \pm 0.00004$, where the uncertainty is 
statistical. A similar calculation in the CDF framework confirms this value.

The uncorrected central value of the D0 electron-channel measurement is 0.23139 
[Eq.~(\ref{eqnD0echan_uncorr})]. To update this measurement to one based on 
templates calculated with NNPDF 3.0 PDFs and \textsc{zfitter}-based electroweak
radiative corrections, the corrections 
$\Delta \sin^2 \theta^{\rm lept}_{\rm eff}({\rm PDF})$, and
$\Delta \sin^2\theta_{\rm eff}^{\rm lept}({\rm \textsc{zfitter}})$ are both applied
for a net correction of $-0.00002 \pm 0.00005$, where the uncertainty
is denoted as ``corrections'' in Table~\ref{d0tblSystErrComb}. 
\begin{table}
\caption{\label{d0tblSystErrComb}
Summary of the D0 systematic uncertainties on the electron-
and muon-channel combination of the electroweak mixing parameter
$\sin^2\theta^{\rm lept}_{\rm eff}$ from
Ref. \cite{D0sw2m10}. The column
labeled $\delta\sin^2\theta^{\rm lept}_{\rm eff}$ gives the
uncertainty of each source.
}
\begin{ruledtabular}
\begin{tabular}{lc}
Source  & $\delta\sin^2\theta^{\rm lept}_{\rm eff}$ \\ \hline
Energy ($e^+e^-$) or momentum ($\mu^+\mu^-$) calibration  & $\pm \: 0.00001$  \\
Energy ($e^+e^-$) or momentum ($\mu^+\mu^-$) resolution   & $\pm \: 0.00002$  \\
Backgrounds                                              & $\pm \: 0.00001$  \\
Charge misidentification                                 & $\pm \: 0.00002$  \\
Lepton identification                                    & $\pm \: 0.00005$  \\
Fiducial asymmetry                                       & $\pm \: 0.00001$  \\
Corrections (PDF and \textsc{zfitter})                   & $\pm \: 0.00005$  \\
NNPDF 2.3 ($e^+e^-$) or NNPDF 3.0 ($\mu^+\mu^-$) PDF      & $\pm \: 0.00019$  \\
\end{tabular}
\end{ruledtabular}
\end{table}
The corrected value is
\begin{eqnarray}
  \sin^2\theta^{\rm lept}_{\rm eff} 
     & = & 0.23137 \pm \: 0.00043 \: {\rm (stat)}  \nonumber \\
     &   & \pm \: 0.00009 \: {\rm (syst)}  \nonumber \\
     &   & \pm \: 0.00017 \: {\rm (PDF)},
\label{eqnD0echan}
\end{eqnarray}
which is the D0 electron-channel value used to combine with the D0
muon-channel and CDF results for the final Tevatron combination.

The published D0 electron-channel value of $\sin^2\theta_{\rm eff}^{\rm lept}$ in
Ref.~\cite{D0sw2e10} includes no PDF correction, and only a partial electroweak
radiative correction that accounts for differences in the effective couplings
to leptons, up-type quarks, and down-type quarks, but not for the complex value
and mass-scale dependence of the couplings. Therefore, the fully-corrected  
Eq.~(\ref{eqnD0echan}) is used for the D0 \cite{D0sw2m10} and Tevatron 
combinations.

The higher-order weak-interaction radiative correction, 
$\Delta \sin^2\theta_{\rm eff}^{\rm lept}({\rm \textsc{zfitter}})$, is applied to 
the uncorrected muon-channel result, Eq.~(\ref{eqnD0mchan_uncorr}),   
giving a $+0.00022$ shift to the measured central value and an additional 
$\pm \: 0.00004$ systematic uncertainty. Applying this correction, the D0 
muon-channel final corrected result \cite{D0sw2m10} is
\begin{eqnarray}
  \sin^2\theta^{\rm lept}_{\rm eff} 
     & = & 0.23016 \pm \: 0.00059 \: {\rm (stat)}  \nonumber \\
     &   & \pm \: 0.00006 \: {\rm (syst)}  \nonumber \\
     &   & \pm \: 0.00024 \: {\rm (PDF)}.
\label{eqnD0mchan}
\end{eqnarray}

The D0 combination result \cite{D0sw2m10} for $\sin^2\theta^{\rm lept}_{\rm eff}$
is obtained using
the corrected electron- and muon-channel results as inputs. The central value 
and systematic uncertainties are combined using the D0 electron- and 
muon-channel central values with the inverse of the squares of the 
measurements' statistical uncertainties as weights. The electron and muon 
systematic uncertainties are treated as uncorrelated, with the exception of the higher-order radiative 
correction uncertainty which is treated as 100\% correlated. However, the total 
uncertainty is marginally affected by this choice, because both the electron- 
and muon-channel total measurement uncertainties are dominated by their 
statistical uncertainties. 
  
The correlation of PDF uncertainties between the electron- and muon-channel 
acceptances cannot be ignored. Instead of estimating the correlation matrix
between the electron and muon channels, we directly estimate the combined PDF
uncertainty. We first estimate the PDF uncertainty on the $A_{\rm{fb}}$ observable
averaged over the electron and muon channels.  We then scale that uncertainty
using the linear relation between $A_{\rm{fb}}$ and $\sin^2\theta_{\rm W}$ 
estimated in simulated events.
  
The D0 combination of electron- and muon-channel results \cite{D0sw2m10} is
\begin{eqnarray}
  \sin^2\theta^{\rm lept}_{\rm eff} 
     & = & 0.23095 \pm \: 0.00035 \: {\rm (stat)}  \nonumber \\
     &   & \pm \: 0.00007 \: {\rm (syst)}  \nonumber \\
     &   & \pm \: 0.00019 \: {\rm (PDF)},
\label{eqnD0comb}
\end{eqnarray}
and a summary of the systematic uncertainties of the combination is presented 
in Table~\ref{d0tblSystErrComb}.

\section{\label{sec:d0cdfCombination}
CDF and D0 combination}

The Tevatron combination of the $\sin^2\theta^{\rm lept}_{\rm eff}$ 
parameter uses a single value from each
experiment, the combined result of the electron- and muon-channel
analyses. The analyses used to derive these values are described in
Sec.~\ref{ssec:CDFsummary} for CDF, and Sec.~\ref{ssec:D0summary}
for D0. Inferences of the $\sin^2\theta_W$ $(M_W)$ parameter
corresponding to the Tevatron-combination value of
$\sin^2\theta^{\rm lept}_{\rm eff}$ and the CDF and D0
input values are obtained using SM calculations in the on-shell
renormalization scheme.

\subsection{\label{ssec:finalCombo}
Tevatron combination}

The central values of the combined electron- and muon-channel results
from each experiment are
\begin{eqnarray}
  \sin^2\theta^{\rm lept}_{\rm eff} = 0.23221 \pm 0.00043  \: ({\rm CDF}),  \; {\rm and} \\
  \sin^2\theta^{\rm lept}_{\rm eff} = 0.23095 \pm 0.00035  \: ({\rm D0}),
\end{eqnarray}
where the uncertainties are statistical only.
The systematic uncertainites are summarized in
Tables~\ref{cdftblSystErrComb} and \ref{d0tblSystErrComb}.
D0 avoids a QCD scale uncertainty by incorporating NNLO effects into the 
$A_{\rm{fb}}$ templates, while CDF avoids sensitivity to lepton identification 
and detector asymmetry uncertainties through the use of the event-weighting 
method described in Sec.~\ref{ssec:CDFsummary}.
The PDF uncertainties are treated as 100\% correlated.
All other systematic uncertainties in Tables~\ref{cdftblSystErrComb} and
\ref{d0tblSystErrComb}
are treated as uncorrelated for the CDF and D0 combination. There are
some correlations from the common \textsc{pythia}-derived backgrounds,
but since the overall contribution from the background is small
and the detectors are different, the backgrounds are treated as
uncorrelated.

The combination of the two inputs with the
``best linear unbiased estimate'' (BLUE) method \cite{BLUE}
yields the Tevatron combination,
\begin{eqnarray}
  \sin^2\theta^{\rm lept}_{\rm eff} 
     & = & 0.23148 \pm \: 0.00027 \: {\rm (stat)} \nonumber  \\
     &   & \pm \: 0.00005 \: {\rm (syst)} \nonumber  \\
     &   & \pm \: 0.00018 \: {\rm (PDF)}.
\label{eqnTevcomb}
\end{eqnarray}
The total uncertainty is $\pm 0.00033$, with
Table~\ref{cdfd0tblErrComb} summarizing the input-source
uncertainties and the corresponding combination
uncertainties. 
\begin{table}
\caption{\label{cdfd0tblErrComb}
Summary of the uncertainties for the electroweak mixing parameter
$\sin^2\theta^{\rm lept}_{\rm eff}$
from the Tevatron combination of the CDF and D0 measurements.
}
\begin{ruledtabular}
\begin{tabular}{lccc}
        & \multicolumn{3}{c}{Uncertainties on $\sin^2\theta^{\rm lept}_{\rm eff}$} \\ \cline{2-4}
        &            &           & Tevatron \\
Source  & CDF inputs & D0 inputs & combination \\ \hline
Statistics          & $\pm \: 0.00043$ & $\pm \: 0.00035$ & $\pm \: 0.00027$  \\
Uncorrelated syst.  & $\pm \: 0.00007$ & $\pm \: 0.00007$ & $\pm \: 0.00005$  \\
PDF                 & $\pm \: 0.00016$ & $\pm \: 0.00019$ & $\pm \: 0.00018$  \\
\end{tabular}
\end{ruledtabular}
\end{table}
The combination weights for the CDF and D0
inputs are 0.42 and 0.58, respectively, and the
$\chi^{2}$ probability for the compatibility of the two
input measurements is 2.6\%.

\subsection{Inference of $\sin^2\theta_W$}
\label{ssec:sw2Inference}

The determination of $\sin^2\theta^{\rm lept}_{\rm eff}$ is a direct
measurement in that the observed asymmetry is directly sensitive to the
effective mixing parameters. In order to obtain $\sin^2\theta_W$ and its
uncertainty, the relationship
\begin{eqnarray}
 \sin^2 \theta^{\rm lept}_{\rm eff} =
   \operatorname{Re}[\kappa_e(\sin^2 \theta_W,M_Z^2)] \sin^2 \theta_W \: 
\label{eqnSin2thetaleff}
\end{eqnarray}
and the \textsc{zfitter} SM calculation of the form factor
$\kappa_e$ with a set of input parameters are required.
The calculation and parameters
specified in the Appendix provide the context for the inference
of $\sin^2\theta_W$. The calculated value of the form factor, 
$\operatorname{Re}[\kappa_e]$, is 1.0371 for the value of the
effective-leptonic mixing parameter specified in Eq.~(\ref{eqnTevcomb})
\cite{zA4ee21prd}.
The choice of the SM-input parameter value for the top-quark mass
affects the value of the form factor, and thus the
inference of $\sin^2\theta_W$. The uncertainty of the inferred
value of $\sin^2\theta_W$ due to the uncertainty from the top-quark
mass input, $173.2 \pm 0.9$~GeV/$c^2$ \cite{topMassCDFD0},
is $\pm 0.00008$ \cite{cdfAfb9eeprd}. This uncertainty is denoted as the 
``form factor'' uncertainty in Table~\ref{sw2ErrComb}.
\begin{table}
\caption{\label{sw2ErrComb}
Summary of uncertainties on the inference of the on-shell
electroweak mixing parameter $\sin^2\theta_W$ for the
Tevatron-combination value of $\sin^2\theta^{\rm lept}_{\rm eff}$.
The column labeled $\delta\sin^2\theta_W$ gives the uncertainty
of each source.
Except for the uncertainty due to the sample size,
all other entries are systematic uncertainties.
}
\begin{ruledtabular}
\begin{tabular}{lc}
Source  & $\delta\sin^2\theta_W$ \\ \hline
Statistics            & $\pm \: 0.00026$  \\
Uncorrelated          & $\pm \: 0.00005$  \\
PDF                   & $\pm \: 0.00017$  \\
Form factor $(m_t = 173.2 \pm 0.9 \; {\rm GeV}/c^2)$
		      & $\pm \: 0.00008$  \\
\end{tabular}
\end{ruledtabular}
\end{table}

For the CDF measurement, the values of
$\sin^2\theta^{\rm lept}_{\rm eff}$, $\sin^2\theta_W$, and $M_W$
based on the combined electron- and muon-channel results are
\begin{eqnarray}
  \sin^2 \theta^{\rm lept}_{\rm eff} = 0.23221 \pm 0.00043 \pm 0.00018  \\
  \sin^2 \theta_W = 0.22400 \pm 0.00041 \pm 0.00019  \\
  M_W = 80.328 \pm 0.021 \pm 0.010 \; {\rm GeV}/c^2 \:,
\label{eqnCDFeffWM}
\end{eqnarray}
and the D0 values of
$\sin^2\theta^{\rm lept}_{\rm eff}$, $\sin^2\theta_W$, and $M_W$
based on the combined electron- and muon-channel results are
\begin{eqnarray}
  \sin^2 \theta^{\rm lept}_{\rm eff} = 0.23095 \pm 0.00035 \pm 0.00020  \\
  \sin^2 \theta_W = 0.22269 \pm 0.00034 \pm 0.00021  \\
  M_W = 80.396 \pm 0.017 \pm 0.011 \; {\rm GeV}/c^2 \:,
\label{eqnD0effWM}
\end{eqnarray}
where the first contribution to each uncertainty is statistical
and the second is systematic. All systematic uncertainties are
combined in quadrature. For $\sin^2 \theta_W$ ($M_W$), the
systematic uncertainties include those propagated from
$\sin^2\theta^{\rm lept}_{\rm eff}$ and the form-factor uncertainty.
\par
The Tevatron-combination values for 
$\sin^2\theta^{\rm lept}_{\rm eff}$, $\sin^2\theta_W$, and $M_W$ are 
\begin{eqnarray}
  \sin^2 \theta^{\rm lept}_{\rm eff} & = & 
           0.23148 \pm 0.00027 \pm 0.00018 \nonumber \\
     & = & 0.23148 \pm 0.00033  \\
  \sin^2 \theta_W  & = &
           0.22324 \pm 0.00026 \pm 0.00019 \nonumber \\
     & = & 0.22324 \pm 0.00033 \\
  M_W & = &
          80.367 \pm 0.014 \pm 0.010 \; {\rm GeV}/c^2 \nonumber \\
    & = & 80.367 \pm 0.017 \;{\rm GeV}/c^2 \:,
\label{eqnCDFD0effWM}
\end{eqnarray}
where the first contribution to each uncertainty is statistical
and the second is systematic. The total systematic uncertainty is
the sum in quadrature of all systematic uncertainties listed in
Tables~\ref{cdfd0tblErrComb} and \ref{sw2ErrComb}.
The form-factor uncertainty is only included in the systematic
uncertainty of $\sin^2 \theta_W$ and $M_W$.

\subsection{\label{ssec:resultComparisons}
Result comparisons}

\begin{figure}
\includegraphics
   [scale=0.4]
   {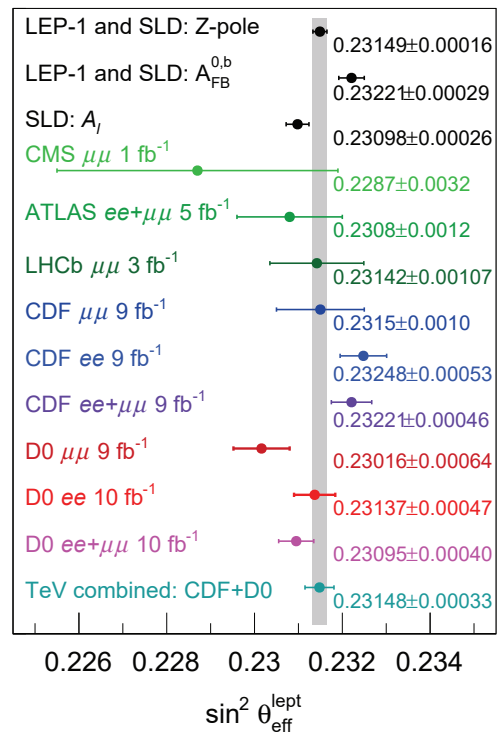}
\caption{\label{fig_compareSW2leff}
Comparison of experimental measurements of 
$\sin^2\theta^{\rm lept}_{\rm eff}$ in the region of the $Z$-boson
pole mass. The horizontal bars represent total uncertainties.
The Tevatron combination (this paper) of CDF and D0 results is denoted
as ``TeV combined: CDF+D0''.
The other measurements are from
LEP-1 combination \cite{LEPfinalZ}, SLD \cite{LEPfinalZ},
CMS \cite{CMSsw2eff1},
ATLAS \cite{ATLASsw2eff},
LHCb \cite{LHCBsw2eff},
CDF \cite{cdfAfb9mmprd,cdfAfb9eeprd}, and
D0 \cite{D0sw2e10,D0sw2m10}.
The LEP-1 and SLD $Z$ pole result is the combination of their six
measurements, and the shaded vertical band shows its uncertainty.
}
\end{figure}

The measurements of $\sin^2\theta^{\rm lept}_{\rm eff}$ are compared 
with previous results from the $Z$-boson pole mass region
in Fig.~\ref{fig_compareSW2leff}.
The hadron-collider results are based on $A_{\rm fb}$ measurements.
The LEP-1 and SLD results
are from the individual asymmetry measurements indicated in the figure.

\begin{figure}
\includegraphics
   [scale=0.35]
   {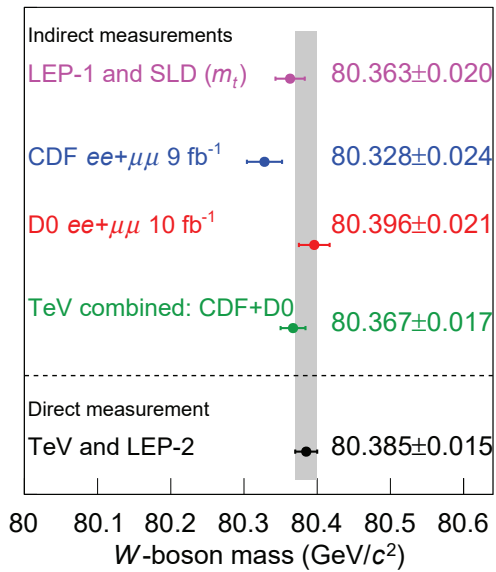}
\caption{\label{fig_compareMW}
Comparison of experimental determinations of the $W$-boson mass 
at high-energy colliders. The horizontal bars represent total uncertainties. 
The Tevatron combination (this paper) based on CDF and D0 results is denoted 
as ``TeV combined: CDF+D0''. 
The other indirect measurements are from 
LEP-1 and SLD \cite{LEPfinalZ,LEPfinalZ2}, 
CDF \cite{cdfAfb9mmprd,cdfAfb9eeprd}, and 
D0 \cite{D0sw2e10,D0sw2m10}. 
All indirect measurements use the Tevatron top-quark mass 
measurement specified in the text \cite{topMassCDFD0}. The SM context for the
Tevatron inferences is specified in the Appendix, and the SM fit of LEP-1 and 
SLD is described in Appendix F of Ref.~\cite{LEPfinalZ2}. For the Tevatron 
inferences of the $W$-boson mass, the SM Higgs-boson mass parameter is fixed, 
while in the LEP-1 and SLD SM fit, it is a floating parameter. The direct 
measurements are from the Tevatron and LEP-2 \cite{tevWmassCDFD0}, and the 
shaded vertical band shows its uncertainty. 
}
\end{figure}

The $W$-boson mass inference is compared in
Fig.~\ref{fig_compareMW} with previous direct and indirect
measurements. The direct measurements are from the Tevatron and
LEP-2 \cite{tevWmassCDFD0}.
The indirect measurements from the Tevatron are derived
from the CDF and D0 measurements of $A_{\rm fb}$, and their
combination.
The indirect measurement of $\sin^2\theta_W$ from LEP-1 and SLD,
$0.22332 \pm 0.00039$, is from a SM fit to all
$Z$-pole measurements \cite{LEPfinalZ,LEPfinalZ2} described in
Appendix~F of Ref.~\cite{LEPfinalZ2}. In that fit, the following
input parameters to \textsc{zfitter} are varied
simultaneously within the constraints of the LEP-1 and SLD data:
the Higgs-boson mass $m_H$, the $Z$-boson mass $M_Z$, the QCD coupling
at the $Z$ pole $\alpha_s(M_Z^2)$, and the QED correction
$\Delta \alpha_{\rm em}^{(5)}(M_Z^2)$. The
top-quark mass $m_t$ is constrained to the value measured directly
at the Tevatron, $173.2 \pm 0.9$ GeV/$c^2$ \cite{topMassCDFD0}.
The precision of the Tevatron indirect measurement almost matches
that of the direct measurement combination from the Tevatron and LEP-2.

\section{\label{sec:theEndSummary}
Summary}

The angular distribution of Drell-Yan lepton pairs provides
information on the electroweak mixing parameter $\sin^2\theta_W$.
The effective-leptonic mixing parameter $\sin^2 \theta^{\rm lept}_{\rm eff}$
is derived from measurements of the forward-backward asymmetry $A_{\rm fb}(M)$
in the polar-angle distribution
by the CDF and D0 experiments, where $M$ is the lepton-pair effective mass.
The measurements are based on the full Tevatron proton-antiproton data sets
collected in 2001-2011.
The CDF measurement is derived from electron and muon pairs from a
$p\bar{p}$ collision sample corresponding to 9~fb$^{-1}$ of integrated
luminosity, and
the D0 measurement is derived from electron and muon pairs from
$p\bar{p}$ samples corresponding to integrated luminosities of
9.7~fb$^{-1}$ and 8.6~fb$^{-1}$ respectively.

The Tevatron combination of the CDF and D0 results yields
\begin{eqnarray}
  \sin^2 \theta^{\rm lept}_{\rm eff} & = & 0.23148 \pm 0.00033.
\label{eqnCDFD0results1}
\end{eqnarray}
The combined result is consistent with LEP-1 and SLD measurements
at the $Z$-boson pole. Based on the SM calculations specified in
the Appendix, the inferences of $\sin^2 \theta_W$ and the $W$-boson
mass are
\begin{eqnarray}
  \sin^2 \theta_W = 0.22324 \pm 0.00033, \; {\rm and} \\ 
  \nonumber \\
  M_W = 80.367 \pm 0.017 \;{\rm GeV}/c^2 \;,
\label{eqnCDFD0results2}
\end{eqnarray}
respectively. 
Within the context of the SM, $\sin^2\theta_W$ and the
$W$-boson mass are related. Comparisons of the indirect
measurements of the $W$-boson mass with those from direct
measurements provide powerful tests of the self-consistency
of the SM.

The combined result on the effective $\sin^2\theta_W$ mixing parameter
at the lepton vertex $\sin^2\theta^{\rm lept}_{\rm eff}$ is the
most precise obtained in hadron collisions. It is consistent
with, and approaches in precision, the best
measurements from electron-positron colliders. 
The values of $\sin^2\theta^{\rm lept}_{\rm eff}$ from hadron
and electron-positron colliders are extracted from a
complementary set of processes. At hadron colliders, the
partonic processes are $q\bar{q} \rightarrow e^+e^-$ and
$\mu^+\mu^-$, and the forward-backward asymmetry is sensitive
to the vertex couplings of the outgoing leptons and the
predominantly light quarks from the hadrons. At electron-positron
colliders, the processes are either leptonic, i.e.
$e^+e^- \rightarrow e^+e^-$, $\mu^+\mu^-$, and $\tau^+\tau^-$,
or mixed, i.e. $e^+e^- \rightarrow q\bar{q}$. While the asymmetry
of a mixed process is analogous to that from hadron collisions,
events with $b$ quarks in the final state yield the best
experimental precision while those with lighter quarks yield
significantly less precision.
The result of the Tevatron combination supports the central
value of $\sin^2\theta^{\rm lept}_{\rm eff}$ derived from the 
LEP-1 and SLD $Z$-pole measurements, and the combined
values from the Tevatron and from LEP-1 and SLD are nearly
identical.

\section*{\label{sec:Acknowledgements}
Acknowldegements}

% acknowledgement_combined.tex             1 August 2017
%
% Acknowledgement paragraph for CDF and D0 in English of August 1, 2017 for APS journals

We thank the staffs at Fermilab and collaborating institutions,
and acknowledge support from the
Department of Energy and the National Science Foundation (United States of America);
the Australian Research Council (Australia);
the National Council for the Development of Science and Technology and
the Carlos Chagas Filho Foundation for the Support of Research in the State of Rio de Janeiro (Brazil);
the Natural Sciences and Engineering Research Council
(Canada);
the Chinese Academy of Sciences and the National Natural Science Foundation of China (China);
the Administrative Department of Science, Technology and Innovation (Colombia);
the Ministry of Education, Youth and Sports (Czech Republic);
the Academy of Finland (Finland);
the Alternative Energies and Atomic Energy Commission and
the National Center for Scientific Research/National Institute of Nuclear and Particle Physics  (France);
the Bundesministerium f\"{u}r Bildung und Forschung (Federal Ministry of Education and Research) and the Deutsche Forschungsgemeinschaft (German Research Foundation) (Germany);
the Department of Atomic Energy and Department of Science and Technology (India);
the Science Foundation Ireland (Ireland);
Istituto Nazionale di Fisica Nucleare (National Institute for Nuclear Physics) (Italy);
the Ministry of Education, Culture, Sports, Science and Technology (Japan);
the Korean World Class University Program and the National Research Foundation (Korea);
the National Council of Science and Technology (Mexico);
the Foundation for Fundamental Research on Matter (The Netherlands);
the National Science Council (Republic of China);
the Ministry of Education and Science of the Russian Federation, 
the National Research Center ``Kurchatov Institute" of the Russian Federation, and the Russian Foundation for Basic Research  (Russia);
the Slovak R\&D Agency (Slovakia);
the Ministry of Science and Innovation and the Consolider-Ingenio 2010 Program (Spain);
the Swedish Research Council (Sweden);
the Swiss National Science Foundation (Switzerland);
the Ministry of Education and Science of Ukraine (Ukraine).
the Science and Technology Facilities Council and The Royal Society (United Kingdom);
the A.P. Sloan Foundation (United States of America);
and the European Union community Marie Curie Fellowship Contract No. 302103.
%

   % input acknowledgement

\section*{\label{sec:appendixZFITTERtitle}
Appendix: ZFITTER}

\appendix
%\appendixheaderoff
\invisiblesection{\label{sec:appendixZFITTER}
ZFITTER}

The effects of virtual electroweak radiative corrections for the
Drell-Yan process are obtained from the $Z$-amplitude form factors
for fermion-pair production according to
$e^+e^- \rightarrow Z \rightarrow f \! \bar{f}$. 
These form factors are calculated by
\textsc{zfitter 6.43}~\cite{Dizet, zfitter621, zfitter642},
which is used with LEP-1, SLD, Tevatron, and LHC  measurement inputs for 
precision tests of the SM \cite{LEPfinalZ,LEPfinalZ2}.

The input parameters to the \textsc{zfitter} radiative-correction
calculation are particle masses, the electromagnetic fine-structure
constant $\alpha_{\rm em}$, the Fermi constant $G_F$, the
strong-interaction coupling
at the $Z$-boson mass $\alpha_s(M_Z^2)$, and the contribution of the
light quarks to the ``running''  $\alpha_{\rm em}$ at the $Z$-boson mass
$\Delta \alpha_{\rm em}^{(5)}(M_Z^2)$.
The scale-dependent couplings are
$\alpha_s(M_Z^2)=0.118 \pm 0.001$~\cite{alphaS09}
and $\Delta \alpha_{\rm em}^{(5)}(M_Z^2)=0.0275 \pm 0.0001$ \cite{alpemh5}.
The mass parameters are
$M_Z = 91.1875 \pm 0.0021$ GeV/$c^2$~\cite{LEPfinalZ,LEPfinalZ2},
$m_t = 173.2 \pm 0.9$ GeV/$c^2$ (top quark) \cite{topMassCDFD0}, and
$m_H = 125$ GeV/$c^2$ (Higgs boson).
Form factors and the $Z$-boson total decay-width $\Gamma_Z$ are
calculated.
The central values of the parameters provide the context of the
\textsc{zfitter} SM calculations.

The \textsc{zfitter} package uses the on-shell renormalization
scheme~\cite{OnShellScheme}, where particle masses are on-shell and
$\sin^2 \theta_W = 1 - M_W^2/M_Z^2$
holds to all orders of perturbation theory by definition.
If both $G_F$ and $m_H$ are specified, $\sin\theta_W$ is not
independent, and is related to $G_F$ and $m_H$ by SM constraints
from radiative corrections. To vary the value of the $\sin\theta_W$
($M_W$) parameter, the value of $G_F$ is not constrained. The $M_W$
value is varied in the range 80.0--80.5 GeV/$c^2$, and
for each value, \textsc{zfitter} calculates $G_F$ and the
form factors. Each set of calculations corresponds to a family
of physics models with SM-like couplings where the parameter
$\sin^2\theta_W$ and the $G_F$ coupling are defined by the
$M_W$ parameter.
The Higgs-boson mass constraint $m_H=125$~GeV/$c^2$ ensures that
the form-factor values remain in the vicinity of SM-fit
values from LEP-1 and SLD~\cite{LEPfinalZ,LEPfinalZ2}.

The form factors are calculated in the massless-fermion
approximation. Consequently, they only depend on the fermion
weak isospin and charge, and are distinguished via three indices,
$e$ (electron type), $u$ (up-quark type), and $d$ (down-quark type).
For the $ee \rightarrow Z \rightarrow q\bar{q}$ process,
the \textsc{zfitter} scattering-amplitude ansatz is
\begin{align}
A_q = &  \: \frac{i}{4} \:
         \frac{\sqrt{2} G_F M_Z^2}
              {\hat{s} - (M_Z^2 - i\,\hat{s} \Gamma_Z/M_Z)} \:
         4T_3^e T_3^q \: \rho_{eq}                   \nonumber \\
      &\times \: [
        \langle \bar{e}| \gamma^\mu (1+\gamma_5) |e \rangle
        \langle \bar{q}| \gamma_\mu (1+\gamma_5) |q \rangle \nonumber \\
      &- \: 4|Q_e| \kappa_e\sin^2 \theta_W \:
         \langle \bar{e}| \gamma^\mu |e \rangle
         \langle \bar{q}| \gamma_\mu (1+\gamma_5) |q \rangle \nonumber \\
      &- \: 4|Q_q| \kappa_q\sin^2 \theta_W \:
         \langle \bar{e}| \gamma^\mu (1+\gamma_5) |e \rangle
         \langle \bar{q}| \gamma_\mu |q \rangle \nonumber \\
      &+ \: 16|Q_e Q_q| \kappa_{eq}\sin^4 \theta_W
         \langle \bar{e}| \gamma^\mu |e \rangle
         \langle \bar{q}| \gamma_\mu |q \rangle ] \:,
\label{eqnScatAmp} 
\end{align}
where $q$ equals $u$ or $d$, the terms $\rho_{eq}$, $\kappa_e$,
$\kappa_q$, and $\kappa_{eq}$
are complex-valued form factors, the bilinear $\gamma$ matrix
terms are covariantly contracted, and
$\frac{1}{2}(1+\gamma_5)$ is the left-handed helicity projector in
the \textsc{zfitter} convention.
The form factors are functions of the $\sin^2\theta_W$
parameter and the Mandelstam $\hat{s}$ variable of the
$e^+e^- \rightarrow Z \rightarrow f \! \bar{f}$ process.
The $\kappa_e$ form factors of the $A_u$ and $A_d$ amplitudes are
not equivalent; however, at $\hat{s} = M_Z^2$, they are numerically
equal.

The $\rho_{eq}$, $\kappa_e$, and $\kappa_q$ form factors can be
incorporated into QCD calculations as corrections to the Born-level
$g_A^f$ and $g_V^f$ couplings,
\begin{eqnarray}
  g_V^f  \rightarrow \sqrt{\rho_{eq}}\,
         T_3^f - 2Q_f \kappa_f \: \sin^2\theta_W ) \; {\rm and} \; g_A^f \rightarrow \sqrt{\rho_{eq}} \, T_3^f , \notag \\
  \label{eqnCorrCoupling}
\end{eqnarray}
where $f = e$ or $q$. The resulting current-current amplitude is
similar to $A_q$, but the $\sin^4 \theta_W$ term contains
$\kappa_e \kappa_q$. This difference is eliminated by adding
the $\sin^4 \theta_W$ term of $A_q$ with the replacement of
$\kappa_{eq}$ with $\kappa_{eq} - \kappa_e \kappa_q$
to the current-current amplitude.
Further details are in Ref. \cite{zA4ee21prd}.

The products $\kappa_f \sin^2 \theta_W$, called
effective mixing terms, are directly accessible from measurements
of the asymmetry in the $\cos \vartheta$ distribution. However,
neither the $\sin^2 \theta_W$ parameter nor the $\hat{s}$-dependent
form factors can be inferred from measurements
without assuming the SM. The effective mixing terms
are denoted as $\sin^2 \theta_{\rm eff}$ to distinguish them from
the on-shell definition of the $\sin^2 \theta_W$ parameter,
$\sin^2\theta_W =  1 - M_W^2/M_Z^2$.
The Drell-Yan process is most sensitive
to the $\sin^2 \theta_{\rm eff}$ mixing term of the lepton vertex,
$\kappa_e \sin^2 \theta_W$. At the $Z$-boson pole, $\kappa_e$ is
independent of the quark flavor, and the flavor-independent value
of $\kappa_e \sin^2 \theta_W$ is commonly denoted as
$\sin^2 \theta^{\rm lept}_{\rm eff}$. 
For comparisons with other measurements, the value of
$\sin^2 \theta^{\rm lept}_{\rm eff}$ at the $Z$-boson pole is taken to be
$\operatorname{Re}[\kappa_e(\sin^2 \theta_W,M_Z^2)] \sin^2 \theta_W$.

\bibliography{sw2TeVcomb_PRD_final.bib}

\end{document}